\documentclass[preprint,aps]{revtex4}

\usepackage{graphicx}

\begin{document}

\title{NUCLEAR RADII OF UNSTABLE NUCLEI}

\author{G. D. Alkhazov}
\affiliation{Petersburg Nuclear Physics Institute, \\
Gatchina, St.Petersburg 188350, Russia}
\author{I. S. Novikov}
\email{ivan.novikov@wku.edu}
\affiliation{Department of Physics and Astronomy, \\
Western Kentucky University, \\
1906 College Heights Blvd, \#11077, Bowling Green, \\
KY 42101-1077, USA}
\author{Yu. Shabelski}
\affiliation{Petersburg Nuclear Physics Institute, \\
Gatchina, St.Petersburg 188350, Russia}

\begin{abstract}

In the presented review we discuss the problem of extraction the parameters of nuclear matter and charge distribution in stable and unstable isotopes. A substantial amount of information on the nuclear radii and other distribution parameters in light exotic nuclei has been obtained from experiments on intermediate-energy nucleus-nucleus collisions. The analyses of these experiments is usually performed with the help of the Glauber Theory.  We review the main assumptions the Glauber Theory and show how this theoretical approach is used to calculate reaction and interaction cross sections. We also show how radii of nuclear matter can be obtained from the analysis of the experimental data on reaction cross sections. In the provided analysis reaction cross sections were calculated in optical and rigid approximations of the Glauber Theory as well as using explicit expressions. Numerical calculations of Glauber's explicit expressions for reaction cross sections were done using Monte Carlo technique. Recent results of the precise measurements of charge radii of light exotic nuclei which were done using the laser-spectroscopy technique as well as the method of the investigation of the nuclear matter distributions in proton elastic scattering experiments in inverse kinematics are also discussed.
\end{abstract}


\maketitle

\renewcommand{\theequation}{\arabic{section}.\arabic{equation}}
\section{\label{sec:Intro}Introduction}

The root-mean-square radii of nuclear matter ($\rm R^m_{rms}$) as well as  nuclear matter and charge distributions contain an important insight on  nuclear potentials and nuclear wave functions. 

It seems that Elton \cite{Elt} was the first to carry out exact calculations of the influence of a finite size nucleus on the elastic scattering of high-energy electrons. He presented calculations of the ratios of differential cross sections for elastic $e$A scattering on extended and point-like nuclei, and showed that in some cases these ratios were very different from unity.

The first detailed information about nuclear charge density distributions  in stable nuclei comes from the experiments of R. Hofst\"adter. Many of his original papers together with other works concerning the discussed topic, including the paper of Elton, can be found in \cite{Hof1}.

The main results of Hofst\"adter and results from other experiments are presented in his review \cite{Hof2}. Various models for charge density distributions $\rho(r)$ were used to describe experimental results. The list includes the point-like model $\rho(r) = \delta(r)$, uniform, Gaussian, exponential and some other distributions. First of all, it was observed that nuclei are characterized by a form factor $S(q)$
\begin{equation}
\label{1.1}
S(q) = \int \rho(r) e^{-iqr} d^3r \;, \; S(0) = 1.
\end{equation}
which decrease rather fast with $q^2$. Therefore, the point-like model can be excluded from further discussion.

Secondly, nuclear matter is not distributed uniformly even in spherical nuclei. The radial dependence of the distribution is somewhat complicated function of the radial variable $r$, the distance from the center of the nucleus. Moreover, these distributions are different in different ranges of atomic weights A and charges Z.   

To describe the size of a nucleus the "effective" nuclear radius  $R_A$ was introduced. The experimental data show that for not very light stable nuclei its "effective" radius depends on atomic weight number A (see, for example, in \cite{Elt1})
\begin{equation}
\label{1.2}
R_A = R_0 A^{1/3} \;, R_0 = 1.2 \div 1.4 {\rm fm}.
\end{equation}
The value of $R_0$ slightly decreases when the atomic weight number A is increasing.

The total inelastic hadron-nucleus cross section can be expressed in a following form:
\begin{equation}
\label{1.3}
\sigma^{(r)}_{hA} = \pi R_A^2.
\end{equation}
The simplest expression for the total inelastic (or reaction) cross section, $\sigma^{(r)}_{AB}$, for interaction between nucleus A (with radius $R_A$) and nucleus B (with radius $R_B$) reads
\begin{equation}
\label{1.4}
\sigma^{(r)}_{AB} = \pi (R_A^{1/3} + R_B ^{1/3})^2.
\end{equation}
More accurate expression was suggested in \cite{BP} by Bradt-Peters
\begin{equation}
\label{1.5}
\sigma^{(r)}_{AB} = \pi R_0^2 (A^{1/3} + B ^{1/3} -c)^2 \;, 
\end{equation}
in which the probability of peripheral collision without nuclear interaction is taken into account. This expression is in reasonable agreement with the experimental data obtained in experiments on interactions of stable nuclei at high energy when $R_0 = 1.48 \pm 0.03$ fm and  $c = 1.32 \pm 0.05$, see \cite{Abd}. However, it is clear that in the case of nuclei with halo the values of these parameters can be significantly different.

Nuclear density distribution can be characterized by a set of its moments
\begin{equation}
\label{1.6}
\langle r^{2n}_A \rangle = \int r^{2n} \rho_A(r) d^3r .
\end{equation}
These moments are suitable when one compares different nuclear density distributions. All odd moments are zero due to obvious symmetry of density distributions.

The zero momentum (n = 0) is equal to unity due to the normalization condition. The second momentum (n = 1) determines the root-mean-square nuclear radius, $R_{\rm rms}$, is defined as
\begin{equation}
\label{1.7}
{\rm R_{rms}} = \sqrt{\langle r^2_A \rangle } \;, \;\;
\langle r^2_A \rangle = \int r^2 \rho_A (r) d^3r \;.
\end{equation}
Generally speaking, the full set of all moments $\langle r^{2n}_A \rangle, n = 1,2,...$ determines the distribution $\rho_A(r)$ for every value of $r$. However, $R_{\rm{rms}}$ by itself is often used to compare nuclei described by different nuclear density distributions (see, for example, \cite{Fesh}). The higher moments are also discussed in the literature (see, for example, \cite{Bush}).  We will use notations $R_m$ and $R_{ch}$ when we consider separately the nuclear matter and nuclear charge r.m.s. radii.

For stable nuclei the information on parameters of nuclear density distribution comes from the data on elastic scattering of fast particles on nuclear targets,~\cite{Elt1}. The comparison of the data on electron and proton elastic scattering gives separate information about proton and neutron distributions in the nucleus,~\cite{CLS,ABV}. The Glauber Theory,~\cite{Gl1,Sit1,Gl2}, is typically used to analyze experiments on interactions with nuclei at energies higher than several hundred MeV. The obtained values of parameters for nuclear matter density and for charge density distributions are presented in~\cite{CLS,ABV,DeDe}. 

In the case of unstable (radioactive) nuclei information comes from the experiments with radioactive beams, since the short lifetime often doesn't allow to make target using those elements. The unstable nuclei are produced through the projectile fragmentation of  primary nuclear beam ($^{11}$B in \cite{Tani}, $^{18}$O in \cite{Al1}, $^{40}$Ar in \cite{Oza1}) on the production target. The experimental production cross  sections for different fragments produced in $^{40}$Ar collisions with Be target can be found in \cite{Oza2}. The produced fragments are then separated by a magnetic analyzing system and rescattered on the nuclear reaction target (Be, C or Al). Again, the Glauber Theory is used to analyze these experiments at energies higher than several hundred MeV.  However, the analytical calculation of all Glauber diagrams for nucleus-nucleus interactions is impossible and some approximate approaches can be used.  

In the presented review we mainly focus on the experimental results on nuclear density distribution for unstable nuclei and on the problems of extracting parameters of their distributions.

In Section~\ref{sec:section2} we discuss the main assumptions of the Glauber Theory and its application to cross section calculations.

The numerical results for the nuclear radii depend on the used approximation of the Glauber Theory. It is illustrated in Section~\ref{sec:section3} for the stable light nuclei, where the applicability of the Glauber theory and the nuclear matter distributions are well established.

Historically, the first experimental study of unstable nuclei at  high energies (790 MeV/nucleon) was presented in \cite{Tani,Tani1}. The interaction cross sections of He, Li and Be isotopes on Be, C, and Al targets were measured. It was found (independently on some analyses problems) that $^{11}$Li has a radius much larger than other neighboring nuclei. This remarkable result suggests the existence of a long tail in the nuclear matter distribution in $^{11}$Li,  i.e. halo. The obtained data stimulated the appearance of many experimental and theoretical papers in which various approaches were used to analyze experimental results obtained in the collisions of unstable nuclear beams with nuclear targets. The analyses of the experimental data and the main results are discussed in Sections~\ref{sec:section4}.

In Section~\ref{sec:section5} we give a short review of the results for electrical charge distributions obtained laser-spectroscopy experiments.

It is necessary to note that experimental measurements of the interaction cross section can provide information on the only one parameter of the nuclear density distribution, or on the only one momentum of the distribution (for example, $R_{\rm rms}$). Different type of experiments in which the differential cross sections for proton elastic scattering $d\sigma/dt$, where $\vert t \vert = q^2$, on unstable nuclei in inverse kinematics are measured, provide better information on the nuclear density distribution and several moments depending on the accuracy of experiment.

In these experiments beam of unstable nuclei is scattered on the proton target and several experimental points at different values of squared transfer momentum $q^2$ can be obtained. That allows to determine separately, for example, the radii of nuclear core and halo. First results obtained using this method at projectile energies of about 700 MeV/nucleon were presented in \cite{Al1}. Data analysis which provided information on core and halo radii was conducted in \cite{Al2} (see Fig.~8). In Section~\ref{sec:section6} we discuss results obtained in the experiments in inverse kinematics. 

Finally, a short summary and conclusion remarks are presented in Section~\ref{sec:section7}.

Due to page limitation of this review we will not discuss the Coulomb contributions to the processes of light ion inelastic collisions. In the case of stable nuclei these contributions are small, about $3 \pm 1$\% \cite{Kob,HL} even for $^{12}$C-Pb interaction cross  section. They can be significantly more important for halo nuclei. For example the analyses of \cite{Kob} estimates them to be $10 \pm 3$\% for $^{11}$Li-Al and $30 \pm 12$\% for $^{11}$Li-Pb collisions. 

\renewcommand{\theequation}{\arabic{section}.\arabic{equation}}

\setcounter{equation}{0}
\section{\label{sec:section2}Glauber Theory for elastic nucleus-nucleus scattering}

\subsection{\label{sec:section2.1}Glauber Theory and its main assumptions}

Glauber Theory \cite{Gl1,Sit1,Gl2} describes interaction between high-energy particles and nuclei. The accuracy of Glauber Theory at intermediate  energies was discussed in detail in \cite{KoKo}. The foundation of the Glauber Theory is the eikonal approximation for fast particle scattering in quantum mechanics. Description of the eikonal approximation can be found in \cite{LaLi}, \cite{LISchiff} or elsewhere. 

Let a fast particle (nucleon) with mass $m$, momentum $k$ and kinetic energy $T$ scatter on the nucleus A, which we treat for the moment as a collection of potential wells of the size $a$ and of the depth $V_0$. Provided the following conditions 
\begin{equation}
\label{2.1}
k a \gg 1 \;, \; T/V_0 \gg 1
\end{equation}
are satisfied, the characteristic scattering angles are small. The phase shift $\chi_A(b)$ is given by the integral of the total nuclear potential (see detailed explanations in \cite{LaLi}, \cite{LISchiff}):
\begin{equation}
\label{2.2}
\chi_A(b) = - \frac{m}{k} \int^{\infty}_{-\infty} dz\, V_A(b,z) \;,
\end{equation}
where $b$ is the impact parameter (2-dimensional vector). Due to the conditions (\ref{2.1}), the incident particle cannot interact more than once with a given target nucleon and the target nucleons have no time for interactions with each other during the scattering process. 

The next important assumption is that the nuclear potential $V_A$ is built up from spatially separated  potentials which are corresponding to nucleons. In this case the phase shift on the nucleus, $\chi_A(b)$, can be represented as a sum of phase shifts for each nucleon-nucleon scattering, $\chi_N(b_i)$: 
\begin{equation}
\label{2.3}
\chi_A(b) = \sum^A_{i=1} \chi_{NN}(b_i) \;.
\end{equation}
In fact we do not have a good parametric characterization for the last  assumption. Indeed, we saw that the distances between the neighboring  nucleons are just of the order of the range of strong interactions. On the  other hand, the small binding energy in comparison with nucleon or pion  mass shows that, as a rule, nucleons within nuclei can be considered as quasi-free.

\subsection{\label{sec:section2.2}Amplitude of elastic Nucleon-Nucleus scattering}

The phases $\chi_A(b)$ and $\chi_{NN}(b_i)$ in Eq.~(\ref{2.3}) are directly related to the elastic nucleon-nucleon scattering amplitudes on the nucleus and the isolated nucleon  respectively:
\begin{equation}
\label{2.4}
f^{el}_{NN}(q) = \frac{ik}{2\pi} \int d^2b \,\Gamma(b) e^{iqb} \;,
\end{equation}

\begin{equation}
\label{2.5}
\Gamma_{NN}(b_i) = 1 - e^{i \chi_{NN}(b_j)} = \frac1{2\pi ik} 
\int d^2q e^{-iqb_i} f^{el}_{NN}(q) \;,
\end{equation}

\begin{equation}
\label{2.6}
\Gamma_A(b;r_1,...,r_A) = 1 - e^{i \chi_A(b;r_1,...,r_A)} = 1 - e^{i \sum^A_{j=1} \chi_j(b-b_i)} \;,
\end{equation}

where $r_1,...,r_A$ are the positions of the nucleons and $b_i$ are their transverse coordinates. Because of condition $\sum_{i=1}^A r_i = 0$, only $A-1$ of the nucleon coordinates are independent. This leads to the so-called center-of-mass motion correction.

Let us now consider the elastic nucleon-nucleus scattering with the transition of the ground state of the target nucleus A to itself,
\begin{equation}
\label{2.7}
F^{el}_{hA}(q) = \frac{ik}{2\pi} \int d^2b e^{iqb} \langle A \vert \Gamma_A(b;r_1,...r_A) \vert A \rangle \;.
\end{equation}

Using Eq. (\ref{2.6}), we obtain
\begin{eqnarray}
\label{2.8}
F^{el}_{NA}(q) & = & \frac{ik}{2\pi} \int d^2b e^{iqb} \int d^3r_1...d^3r_A \cdot \rho(r_1,...,r_A) \times  \\ \nonumber
& \times & \left\{1 - \prod_{i=1}^A \left[1 - \frac{1}{2\pi ik} \int d^2q_1 e^{-iq_1(b-b_i)} f_{NN}(q_1) \right] \right\} \;.
\end{eqnarray}
Here $b_i$ is the transverse coordinate of the $i$-th nucleon, and $\rho(r_1,...,r_A)$ is the probability density distribution to find nucleons with coordinates $(r_1, r_2, ..., r_A)$.

Assuming that the internucleon interactions are rather week, we can neglect the possible internucleon correlations\footnote{Such correlations can be included in some models, see, for example, \cite{ABV}.}. In this case probability density distribution $\rho(r_1,...,r_A)$ can be reduced to the product of one-particle densities $\rho(r_i)$ 
\begin{equation}
\label{2.9}
\rho(r_1,...,r_A) = \prod^A_{i=1} \rho(r_i) \;, 
\;\; \int d^3r_i \rho(r_i) =1 \;.
\end{equation}
Using assumption Eq.~(\ref{2.9}) and integrating over the positions of the nucleons we can get the following expression for $F^{el}_{NA}(q)$
\begin{equation}
\label{2.10}
F^{el}_{NA}(q)  =  \frac{ik}{2\pi} \int d^2b e^{iqb} \left[1 - \left(1 - \frac{1}{2\pi ik} \int d^2q_1 e^{-iq_1b} f^{el}_{NN}(q_1) S(q_1) \right)^A \right] \;,
\end{equation}
where $S(q_1)$ is the one-particle nucleus form factor determined by Eq.~(\ref{1.1}).

\subsection{\label{sec:section2.3}Amplitude of elastic Nucleus-Nucleus scattering}

In the framework of the Glauber Theory the elastic scattering amplitude of nucleus A on nucleus B with momentum transfer $q$ can be expressed as \cite{PaRa,BrSh}
\begin{equation}
\label{2.11}
F^{el}_{AB}(q) = \frac{ik}{2\pi} \int d^2b \, e^{iqb} \, \langle A \vert \langle B \vert \Gamma_{AB}(b;r_1,...r_A;r'_1,...r'_B)\vert B \rangle \vert A \rangle  \;.
\end{equation}
Here $k$ is the incident momentum of one nucleon in A-nucleus in laboratory frame, and $b$ is an impact parameter. This expression is written in the frame, where B-nucleus is a fixed target.

Let us assume, similarly to Eq.~(\ref{2.3}), that the phase shift in the nucleus-nucleus scattering, $\chi_{AB}(b)$, is equal to the sum of phase shifts for each nucleon-nucleon scattering, $\chi_{NN}(b_i)$. Integrating over the longitudinal coordinates of all nucleons we obtain
\begin{equation}
\label{2.12}
F^{el}_{AB}(q) = \frac{ik}{2\pi} \int d^2b e^{iqb} [1 - S_{AB}(b)] \;,
\end{equation}
where
\begin{equation}
\label{2.13}
S_{AB}(b) = \langle A \vert \langle B \vert \left\{
\prod_{i, j} [1 - \Gamma_{NN}(b + u_i - s_j)] \right\} \;
\vert B \rangle \vert A \rangle \;,
\end{equation}
where $u_i$ and $s_j$ are the transverse coordinates of nucleons in the nuclei A and B, respectively.

Contrary to the case of nucleon-nucleus interaction, integral in Eq.~(\ref{2.13}) cannot be evaluated analytically even with the  assumption Eq.~(\ref{2.9}) that nuclear  densities $\rho(r_1,...,r_A)$ in both $A$ and $B$ nuclei are the normalized products of one-nucleon densities $\rho(r_i)$.  

To make the problem manageable, one can retain only part of all  contributions  in the expansion of the product in Eq.~(2.13), that corresponding to the  contributions characterized by large combinatorial factors. The leading graphs correspond to the so-called optical approximation, \cite{CzMa,Karol}, in which  one sums up the contributions with no more than one scattering for each nucleon.  In other words, only those products of amplitudes $\Gamma_{NN}(b + u_i - s_j)$   in Eq.~(2.13) are taken into accounts which have different indices $i$, $j$. This approximation corresponds to the summation of diagrams shown in Figs.~1a, 1b,  1c,...  In this approximation diagram which describes $n$-fold interaction has a  combinatorial factor $_{A}C_{n} \; _{B}C_{n}$. To avoid crowding of  lines in Fig.~1 we have only shown the nucleon participants from the nucleus $A$ (upper  dots) and $B$ (lower dots) with the links standing for interacting amplitudes,  and we have not plotted the nucleon-spectators. 

\begin{figure}[th]
\includegraphics[width=0.9\textwidth]{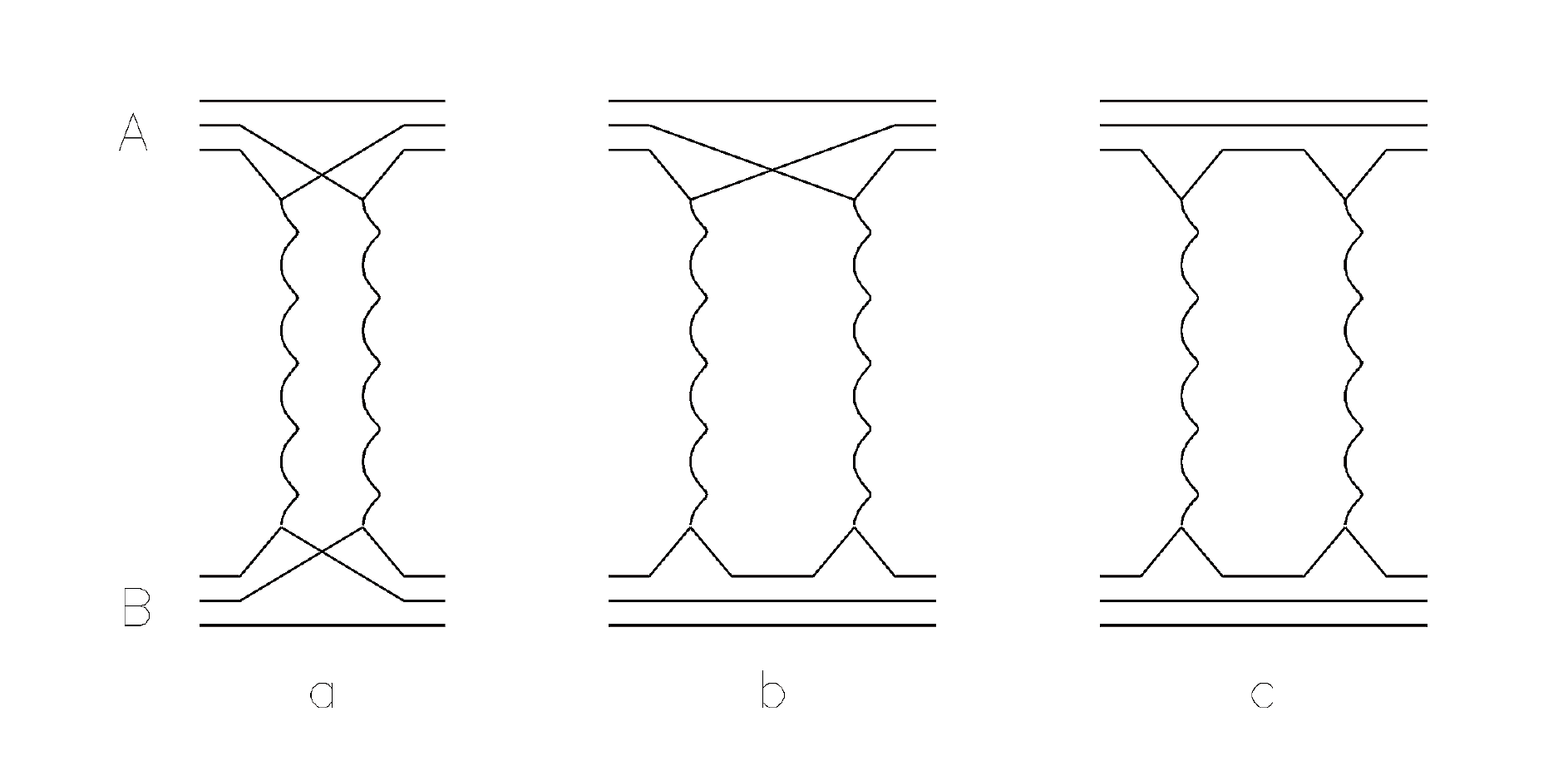}
\caption{Diagrams of the interaction of two nuclei, $A$ and $B$, taken into account.}
\end{figure}

In the optical approximation averaging $\langle A \vert ... \vert A \rangle$ and $\langle B \vert ... \vert B \rangle$ of the product  $\prod_{i,j}  [1 - \Gamma_{NN}(b + u_i - s_j)]$ can be substituted with the averaging of the $[1 - \Gamma_{NN}(b + u_i - s_j)]$
\begin{equation}
\label{2.14}
S_{AB}^{opt}(b) = \prod_{i,j} \langle A \vert \langle B \vert
[1 - \Gamma_{NN}(b + u_i - s_j)] \; \vert B \rangle \vert A \rangle \;.
\end{equation}

The amplitude of elastic nucleon-nucleon scattering can be described by the standard parameterization
\begin{equation}
\label{2.15}
f^{el}_{NN} (q) = \frac{ik\sigma^{tot}_{NN}}{4 \pi} \exp \left( -\frac12 \beta q^2\right).
\end{equation}
Here $\sigma^{tot}_{NN}$ is the total NN cross section, $\beta$ is the slope parameter of the differential NN cross-section dependence 
on $q^2$. We neglect the real part of $f^{el}_{NN} (q)$.

When both nuclei are not very light, using the standard assumptions of the multiple scattering theory, scattering matrix $S_{AB}$ can be written in a following form
\begin{equation}
\label{2.16}
S_{AB}^{opt}(b) = \left [1 - \frac1A T_{opt}(b) \right ]^A \approx
\exp [- T_{opt}(b)] \;,
\end{equation}
where
\begin{equation}
\label{2.17}
T_{opt}(b) = \frac{\sigma^{tot}_{NN}}{4 \pi \beta} \int d^2b_1 d^2b_2 T_A(b_1) T_B(b_2)
\exp\left[-\frac{(b+b_1-b_2)^2}{2\beta}\right]
\end{equation}
with $T_A$ and $T_B$
\begin{equation}
\label{2.18}
T_A(b) = A \int^{\infty}_{-\infty} dz \rho_A\left( r=\sqrt{b^2+z^2} \right) \;.
\end{equation}

Neglecting the NN interaction range $\beta$ in comparison to the nuclear radii, we have
\begin{equation}
\label{2.19}
T_{opt}(b) = \frac{\sigma^{tot}_{NN}}{2} \int d^2b_1 T_A(b-b_1) T_B(b_1) \;.
\end{equation}

In the diagram language Eq. (\ref{2.13}) accounts for all possible intermediate states of nucleons between the interactions, as it is shown in Fig.~2a, while the optical approximation, Eq.~(2.17), would correspond to the interactions with only one pole (nuclear ground state) in the both $A$ and $B$ intermediate states, \cite{BrSh}
(see Fig.~2c). 

\begin{figure}[th]
\includegraphics[width=0.9\textwidth]{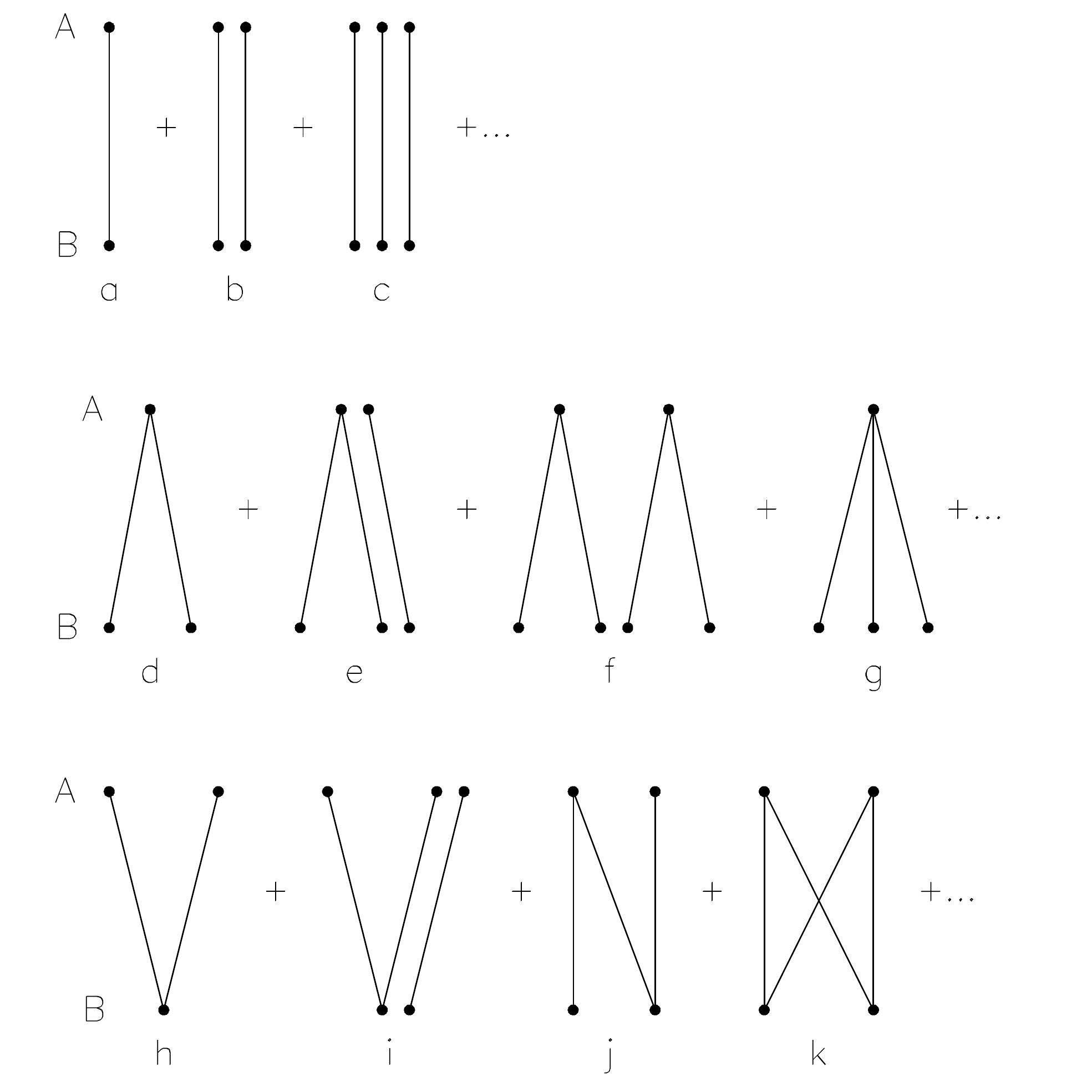}
\caption{Two-fold interaction of two nuclei in the multiple scattering theory (a), in the rigid target approximation (b) and in the optical approximation (c).}
\end{figure}

Unfortunately, numerical calculations in~\cite{FV} (see also~\cite{Sh2} for the case of collisions of very light nuclei) demonstrate that the optical  approximation is not accurate enough even for the integrated cross sections.  The difference with the data amounts for $\sim 10-15$ \% in $\sigma_{AB}^{tot}$  and it is even greater for differential cross sections,~\cite{FV}. This disagreement can be explained by the fact that series with smaller  combinatorial factors in Eq.~(\ref{2.13}) give significant global corrections  to the optical approximation results. As a matter of fact, the terms of the series are alternating in sign, so, due to the cancelations of terms  with opposite signs, the final sums of these series can have very  different values. Thus, some classes of diagrams with non-leading  combinatorial factors give significant contributions to the final total value.

The rigid target (or rigid projectile) approximation, described in \cite{Alk,VT}, is more explicit than the optical approximation. It corresponds to averaging $\langle B \vert ... \vert B \rangle$ inside the product in Eq.~(\ref{2.13}):
\begin{eqnarray}
\label{2.20}
S_{rg}^{AB}(b) = &&  \langle A \vert \left \{ \prod_{i,j} \vert \langle B \vert [1 - \Gamma_{NN}(b + u_i - s_j)] \vert B \rangle \right \} \vert A \rangle = [T_{rg}(b)]^A \;,
\end{eqnarray}
where
\begin{eqnarray}
\label{2.21}
T_{rg}(b) =&&  \frac1A \int d^2b_1 T_A(b_1) \exp \left \{-\frac{\sigma_{NN}^{tot}}{4 \pi \beta} \int d^2b_2 T_B(b_2) \times \right. \nonumber\\
&& \left. \times \exp\left[-\frac{(b+b_1-b_2)^2}{2\beta}\right] \right \} \;.
\end{eqnarray}
Neglecting the NN interaction range in comparison to the nuclear radii $T_{rg}(b)$ can be simplified as
\begin{equation}
\label{2.22}
T_{rg}(b) = \frac1A \int d^2b_1 T_A(b_1 - b) \exp \left [ - \frac{\sigma_{NN}^{tot}}{2} T_B(b_1) \right ] \;.
\end{equation}

This approximation corresponds to the sum of the diagrams in   Figs.~1a, 1b, 1c, ... and the correction diagrams in Figs.~1d, 1e, 1f, 1g,...  Diagrams in Figs.~1d, 1e, 1f, 1g,... represent the case when each nucleon  from the nucleus A can interact several times, but all interacting nucleons  from B are still different. Each correction diagram which describes  $n$-fold interaction has a combinatorial factor smaller than the combinatorial  factor leading diagrams (Fig.~1a, 1b, 1c, ...).  Although, due to the obvious asymmetry in contributions of the two nuclei  such approach can be theoretically justified in the limit $A/B \ll 1$,  i.e. for C-Pb collisions, this approximation can be used sometimes  in the case of heavy ion collisions with equal atomic weights. 

It was suggested in \cite{Abu} to present an expressions for the nucleus-nucleus scattering as a standard Glauber picture, where the amplitude of nucleon-target scattering is considered as an elementary particle and, after that, all Glauber rescatterings are taken in account. In \cite{AlLo} it was shown that such a picture is equivalent to the rigid  target, or to the rigid projectile approach. 

Further corrections to the elastic amplitude (some of them are shown in  Figs.~1h, 1i, 1j, 1k,...) have been considered in \cite{Andr1,Andr2,MBra,BoKa}. However, the results of such corrections  are somewhat complicated for practical use. 

The possibility to obtain the Glauber Theory results without any simplification  by the direct calculation of Eq.~(\ref{2.13}) using Monte Carlo simulation was  first suggested in~\cite{ZUS,Shm}. This method was used for numerical  calculations in~\cite{Sh11,Gar,Var,AlLo}. The simplest algorithm considering  the values of coordinates uniformly distributed in the interaction region  cannot be applied here because in most cases several coordinates have values corresponding to very small nuclear density.

The algorithm proposed by Metropolis {\it et al} in~\cite{Metro} allows to generate a set of coordinates which are distributed according to a pre-defined  distribution. The Metropolis method (often reffered as Metropolis-Hastings~\cite{Hast} algorithm) is as follows: 
\begin{enumerate}
\item The initial coordinate $s_i$ is randomly generated from the appropriate interval
\item To obtain next coordinate, a shift $\Delta s_i$ is randomly generated and then added to the initial coordinate
\item New coordinate is accepted when the ratio $r = \rho(s_i + \Delta s_i)/\rho(s_i) >1$
\item If the ratio $r<1$, new coordinate is accepted only when $r>x$, where $x$ is a new random number $x$ from $[0,1]$ interval. Otherwise new coordinate is not accepted
\end{enumerate}
Generated set of coordinates can be used to calculate average value of $\prod_{ij} [1 - \Gamma_{NN}(b + u_i - s_j)]$ and, $S_{AB}(b)$. Some results of cross section calculations using described technique are discussed in Section~\ref{sec:section2}.

\subsection{\label{sec:section2.4}Difference between reaction and interaction cross sections for light ion collisions} 

The total inelastic (reaction) cross section for the collisions of nuclei A and B, $\sigma^{(r)}_{AB}$, is equal to the difference of the total interaction cross section $\sigma^{tot}_{AB}$ 
\begin{equation}
\label{2.23}
\sigma^{tot}_{AB} = \frac{4 \pi}k \mathrm{Im} \, F^{el}_{AB}(q=0) = 2 \int d^2b [1 - S_{AB}(b)]
\end{equation}
and integrated elastic scattering cross section $\sigma^{el}_{AB}$:
\begin{equation}
\label{2.24}
\sigma^{el}_{AB} =  \int d^2b [1 - S_{AB}(b)]^2 \;.
\end{equation}
So, for the reaction cross section we obtain
\begin{equation}
\label{2.25}
\sigma^{(r)}_{AB} = \sigma^{tot}_{AB} - \sigma^{el}_{AB} = \int d^2b [1 - \vert S_{AB}(b)\vert^2] \;
\end{equation}
The values of $S_{AB}(b)$ in these expressions can be calculated in one of the approaches described in Section~\ref{sec:section3}.

As it was mentioned in Introduction, in the high energy light ion scattering experiments only interaction cross sections $\sigma^{(I)}_{AB}$ are measured. The difference between interaction and reaction cross sections is that the reaction  cross sections include the cross sections of all processes except of the  elastic scattering AB $\to$ AB, whereas the interaction cross sections do  not include the elastic scattering AB $\to$ AB as well as the processes  with a target nuclei excitation or disintegration AB $\to$ AB$^*$ ($B^* \neq B$), so $\sigma^{(I)}_{AB} < \sigma^{(r)}_{AB}$. The difference between  $\sigma^{(I)}_{AB}$ and $\sigma^{(r)}_{AB}$ was estimated in  \cite{KSh,Oga} to be not larger than a few percents of their values for a beam energy higher than several hundred MeV per nucleon. Usually this difference is neglected in the analyses of experimental data, see,  for example, \cite{AlLo,Oza,MNS}. 

Lets define A$^\prime$ and B$^\prime$ as all excitation or disintegration states of A and B nucleons including their ground states, and A$^*$ and B$^*$ as all excitation or disintegration states excluding their ground states. Therefore, difference between  $\sigma^{(I)}_{AB}$ and $\sigma^{(r)}_{AB}$ can be expressed as $\sigma^{(r)}_{AB} - \sigma^{(I)}_{AB} = \sigma(AB \to AB^*)$. 

The cross sections of the  processes AB $\to$ AB$^*$, as well as the processes AB $\to$ A$^*$B$^*$, where both nuclei can be exited or disintegrated, can be calculated within the Glauber Theory using the same assumptions as in Section~\ref{sec:section2.1}.

Let us first calculate the cross section of the processes AB $\to$ A$^*$B$^*$. The processes when an incident nucleus is excited\footnote{Many of unstable nuclei, for example $^{11}$Li \cite{Tani2} have no excitation levels.} without changing its Z and N numbers are also very useful for our analysis of the difference between reaction and interaction cross sections, since they are not included into $\sigma^{(I)}_{AB}$ cross section. 
In the case of AB $\to$ A$^\prime$B$^\prime$, the amplitude of AB scattering together with the excitation or disintegration of one or both nuclei can be written in a form similar to Eq.~(\ref{2.11})
\begin{equation}
\label{2.26}
F_{AB \to A^\prime B^\prime}(q) = \frac{ik}{2\pi} \int d^2b \, e^{iqb} \, \langle A \vert 
\langle B \vert \Gamma_{AB}(b;r_1,...r_A;r'_1,...r'_B)
\vert B^\prime \rangle \vert A^\prime \rangle  \;.
\end{equation}
Since all processes of $AB \to A^\prime B^\prime$ transitions are the results  of elastic NN scattering the operator $\Gamma_{AB}(b;r_1,...r_A;r'_1,...r'_B)$ has the same form as in Eq.~(\ref{2.11}). 

The difference between the total cross section $\sigma^{tot}_{AB}$ and $\sigma( AB \to A^\prime B^\prime)$ determines the cross  section of secondary particle (pion) production, $\sigma^{prod}_{AB}$. Here we neglect the contributions of the processes where a pion can be produced  in one NN interaction and can be absorbed by another nucleon or nucleon pair.

The cross section of all processes  AB $\to$ A$^\prime$ B$^\prime$
\begin{equation}
\label{2.27}
\frac{d\sigma_{AB \to A^\prime B^\prime}}{d^2q} = \frac1{k^2} \sum_{A^\prime,B^\prime} 
\vert F_{AB \to A^\prime B^\prime}(q) \vert^2
\end{equation}
can be calculated, following method described in \cite{FG}. 

Using the completeness condition for nuclear wave functions
\begin{equation}
\label{2.28}
\sum_{A^\prime, B^\prime} \vert A^\prime B^\prime \rangle \langle A^\prime B^\prime \vert = \prod_{i,j} \delta(u_i-u'_i) \delta(s_j-s'_j) \;,
\end{equation}
the cross section of AB $\to$ A$^\prime$ B$^\prime$ can be written as
\begin{equation}
\label{2.29}
\sigma_{AB \to A^\prime B^\prime} = \int [1 - 2S_{AB}(b) +  I_{AB}(b)] d^2b \;,
\end{equation}
where
\begin{equation}
\label{2.30}
I_{AB}(b) = \langle A \vert  \langle B \vert \left\{ \prod_{i,j} [1 - \Gamma_{NN} (b + u_i - s_j)] \right\}^2 \vert B \rangle \vert A \rangle \;.
\end{equation}

The cross section of the processes AB $\to$ A$^*$B$^*$, i.e. processes with the excitation or disintegration of one or both nuclei without the elastic scattering channel, can be calculated as $\sigma (AB \to A^*B^*) = \sigma (AB \to A^\prime B^\prime) - \sigma (AB \to AB)$. Therefore, $\sigma (AB \to A^*B^*)$ can be expressed in the following form
\begin{equation}
\label{2.31}
\sigma_{AB \to A^*B^*} = \int [I_{AB}(b) - S^2_{AB}(b)] d^2b \;. 
\end{equation}

$I_{AB} (b)$ can be simplified using parameterization of elastic scattering nucle\-on-nucleon amplitude Eq.~(\ref{2.15}) and the optical approximation 
\begin{equation}
\label{2.32}
I^{opt}_{AB}(b) =  \exp \left( -T^{**}_{opt}(b) \right) \, ,
\end{equation}
where
\begin{eqnarray}
\label{2.33}
T^{**}_{opt}(b)= \frac1{2\pi\beta}\int d^2b_1 d^2b_2 && T_A(b_1) T_B(b_2)\left(\sigma^{tot}_{NN} e^{-\frac{(b+b_1-b_2)^2}{2\beta}}-\right.\nonumber\\
&& \left. -2 \sigma^{el}_{NN} e^{-\frac{(b+b_1-b_2)^2}{\beta}}\right)
\end{eqnarray}

When the radius of NN interaction can be neglected the expression in Eq.~(\ref{2.33}) can be simplified even more \cite{BrSh}
\begin{equation}
\label{2.34}
I^{opt}_{AB}(b) =  \exp \left[-\sigma^{in}_{NN} \int d^2b T_A(b_1) T_B(b-b_1) 
\right] \;,
\end{equation}
where $\sigma_{NN}^{in}$ is the nucleon-nucleon inelastic cross section. 

As we mentioned above the difference between reaction and interaction cross sections is defined by the $\sigma(AB \to AB^*)$. It is easy to see that $\sigma(AB \to AB^*) = \sigma(AB \to AB^\prime) -\sigma(AB \to AB)$, if the  projectile nucleus A has no excitation states. The amplitude of AB $\to$ AB$^\prime$  processes has the form
\begin{equation}
\label{2.35}
F_{AB \to AB^\prime}(q) = \frac{ik}{2\pi} \int d^2b \, e^{iqb} \, \langle A \vert \langle B \vert \Gamma_{AB}(b;r_1,...r_A;r'_1,...r'_B)
\vert B^\prime \rangle \vert A \rangle  \;.
\end{equation}

Again, using the completeness relation for the target B
\begin{equation}
\label{2.36}
\sum_{B^\prime} \vert AB^\prime \rangle \langle AB^\prime \vert = \prod_{j} \vert A \rangle \langle A \vert \, \delta(s_j-s'_j) \;,
\end{equation}
we obtain
\begin{equation}
\label{2.37}
\sigma_{AB \to AB^\prime} = \int [1 - 2S_{AB}(b) +  J_{AB}(b)] d^2b \;,
\end{equation}
where
\begin{eqnarray}
\label{2.38}
J_{AB}(b) = &&\langle A \vert  \langle B \vert \left\{ \prod_{i,j} [1 - \Gamma_{NN} (b + u_i - s_j)] \right\}\vert A \rangle \times \nonumber\\
&& \times \langle A \vert \left\{ \prod_{i,j'} [1 - \Gamma_{NN} (b + u_i - s_j')] \right\}\vert B \rangle \vert A \rangle \;.
\end{eqnarray}

Using expression for the elastic cross section Eq.~(\ref{2.24}) we obtain following expression for $\sigma(AB \to AB^*)$ 
\begin{equation}
\label{2.39}
\sigma_{AB \to AB^*} = \int [J_{AB}(b) - S^2_{AB}(b)] d^2b  
\end{equation}

In the optical approximation $J_{AB}$ can be expressed as
\begin{equation}
\label{2.40}
J^{opt}_{AB} =  \exp(-T^{*}_{opt}(b)) \, ,
\end{equation}
where
\begin{eqnarray}
\label{2.41}
T^{*}_{opt}(b) = \frac{\sigma^{tot}_{NN}}{2\pi\beta} && \int d^2b_1 d^2b_2 T_A(b_1) T_B(b_2) e^{-\frac{(b+b_1-b_2)^2}{2\beta}} \times \nonumber\\
&& \times \left(1 - 2 \frac{\sigma^{el}_{NN}}{\sigma^{tot}_{NN}} \frac1B \int d^2b_3 T_B(b_3) e^{-\frac{(b+b_1-b_3)^2}{2\beta}}\right) \;.
\end{eqnarray}
As usual we are using the parameterization of elastic scattering nucle\-on-nucleon amplitude Eq.~(\ref{2.15}).

If the radius of NN interaction can be neglected $T^*_{opt}$ can be simplified 
\begin{equation}
\label{2.42}
T^{*}_{opt}(b) =  \sigma^{tot}_{NN} \int d^2b T_A(b_1) T_B(b-b_1) \left( 1 - \frac{\sigma^{tot}_{NN}}{4A} T_B(b-b_1)\right) \;.
\end{equation}

The difference between $\sigma^{(r)}_{AB}$ and $\sigma^{(I)}_{AB}$ for radioactive isotope $^{34}$Cl and $^{12}$C target was estimated experimentally in \cite{Oza1}. It was found to be approximately 10 mb (i.e. about 1\% of $\sigma^{(r)}_{AB}$) and in agreement with the calculations presented in \cite{Oga}. The analyses in \cite{KIO} results in larger difference between  $\sigma^{(r)}_{AB}$ and $\sigma^{(I)}_{AB}$. It was found that for experiments where projectiles with with A $<$ 80 interacted with a carbon target the average difference between interaction and reaction cross sections was approximately 60 mb, i.e. about 4-6\% of $\sigma^{(r)}_{AB}$. However, provided in \cite{KIO} results are not accurate (especially for the light nuclei, see in \cite{KIO1}), since the "black-sphere" model with rectangular distribution  of nuclear matter was used for analyses

\subsection{\label{sec:section2.5}One or several nucleon removal cross sections for stable nuclei}

Glauber Theory allows to calculate cross sections of one or several nucleons removal from the projectile nucleus A in the process of nucleus-nucleus collision.

As we mentioned above, assuming that the internucleon interactions are rather week, we can neglect the possible internucleon correlations and express the probability density distribution $\rho(r_1,...,r_A)$ as a product of one-particle density distributions $\rho(r_i)$ (see Eq.~(\ref{2.9})). In this case the cross sections of one or several nucleons removal can be calculated using the AGK \cite{AGK,Shab} cutting rules technique for all Glauber diagrams shown in Fig.~1 and to consider all possible intermediate states. As a result, the cross  sections of removal of one, two, three, etc. nucleons $\sigma^{(1)}_{AB}$, $\sigma^{(2)}_{AB}$, $\sigma^{(3)}_{AB}$, etc. in the arbitrary AB collisions have  the forms \cite{ShTr}  

\begin{equation}
\label{2.43}
\sigma^{(1)}_{AB} = A (\sigma^{(r)}_{AB} - \sigma^{(r)}_{A-1, B}) \;,
\end{equation}
\begin{equation}
\label{2.44}
\sigma^{(2)}_{AB} = \frac{A(A-1)}{2!} (-\sigma^{(r)}_{AB} + 2\sigma^{(r)}_{A-1, B} -2\sigma^{(r)}_{A-2,B}) \;,
\end{equation}
\begin{equation}
\label{2.45}
\sigma^{(3)}_{AB} = \frac{A(A-1)(A-2)}{3!} (\sigma^{(r)}_{AB} - 3\sigma^{(r)}_{A-1, B} + 3\sigma^{(r)}_{A-2,B} - \sigma^{(r)}_{A-3,B}) \;.
\end{equation}
Here we took into account all Glauber diagrams shown in Fig.~1, including loop diagram Fig.~1k without any multidimensional integration. 

An important point is that the cross sections $\sigma^{(r)}_{A-1,B}$, $\sigma^{(r)}_{A-2,B}$, $\sigma^{(r)}_{A-3,B}$  in Eqs. (\ref{2.43})-(\ref{2.45}) cannot be taken from the experimental data.  The cross sections in Eqs. (\ref{2.43})-(\ref{2.45}) are results of the results of cancellation of the diagrams contributions (diagrams are shown in Fig.~1b, 1e, 1h, etc), which contain the radii of nuclei with weights A, A-1, A-2 in denominators. Therefore, cross sections $\sigma^{(r)}_{A-1,B}$, $\sigma^{(r)}_{A-2,B}$, $\sigma^{(r)}_{A-3,B}$ should be calculated in the framework of the Glauber Theory with the same nuclear radii, as  $\sigma^{(r)}_{AB}$.

The equations (\ref{2.43})-(\ref{2.45}) are exact in the framework of the Glauber  Theory. However, their application for the calculation of the physical processes,  for example, to the cross sections of removal of several nucleons with the condition  that the nuclear remnant stay to be bound needs in some additional physical assumptions.  

Really, we assume that in the process of NN scattering with transfer momentum  of the order of that in free NN interaction both nucleons will be removed  from the nuclei but the nuclear remnants will stay bound. It seems to be true  for the collisions of light nuclei with an accuracy about 80\% \cite{Gla}   

The case of the fragmentation of unstable nuclei is discussed below in Section~\ref{sec:section5.3}.

\setcounter{equation}{0}
\section{\label{sec:section3}Monte Carlo simulation of reaction nucleus-nuc\-le\-us cross sections}

\subsection{\label{sec:section3.1}Parameters of NN amplitude}

In this section we will discuss numerical method to calculate reaction cross section $\sigma_{AB}^{(r)}$. Below we present the results of the analysis the data obtained in the at high energy nucleus-nucleus collision experiments at energies about 800 MeV per nucleon. The parameters of NN elastic scattering amplitude at these energies were analyzed in \cite{Shab1}. At energy 800 MeV per nucleon we obtained
\begin{eqnarray}
\label{3.1}
\beta^2 && = 5 \; {\rm GeV}^{-2} = 0.2 \; {\rm fermi}^2 \nonumber\\
\sigma^{tot}_{pp} &&= 47 \; {\rm mb};  \; \sigma^{tot}_{pn} = 38 \; {\rm mb}\\
\sigma^{el}_{pp} &&= 25.4 \; {\rm mb}; \; \sigma^{el}_{pn} = 26 \; {\rm mb}  \nonumber
\end{eqnarray}

For the comparison and possible interpolation we present the same quantities at energy 1 GeV per nucleon: 
\begin{eqnarray}
\label{3.2}
\beta^2 && = 5.5 \; {\rm GeV}^{-2} = 0.22 \; {\rm fermi}^2 \nonumber\\
\sigma^{tot}_{pp} &&= 47.5 \; {\rm mb};  \; \sigma^{tot}_{pn} = 38.2 \; {\rm mb}\\
\sigma^{el}_{pp} &&= 24.65 \; {\rm mb}; \; \sigma^{el}_{pn} = 23.956 \; {\rm mb}  \nonumber
\end{eqnarray}

The ratio of the real to imaginary part of elastic scattering amplitude for p$^4$He was obtained to be equal $0.06 \pm 0.06$ at energy 700 MeV, \cite{Alk1a}. Since the term proportional to the second dergee of this ratio contributes to light ion cross sections, so it can be neglected in the further calculations.  

\subsection{\label{sec:section3.2}Extraction of ${\rm R_m}$ values in various theoretical approaches} 

As we mentioned in Introduction, for stable nuclei the information on parameters of nuclear density distribution comes from the data on elastic scattering of fast particles on nuclear targets. 

The total inelastic (reaction) cross section for the collisions of nuclei A and B, $\sigma^{(r)}_{AB}$ is shown in Eq. (\ref{2.23}):
\[
\sigma^{(r)}_{AB} = \int d^2b [1 - \vert S_{AB}(b)\vert^2] \;.
\]

For the numerical calculations of $S_{AB}$ it is necessary to use an expression for the  nuclear matter density distributions in the colliding nuclei. The most detailed information about these distributions comes from the data on electron or hadron differential elastic scattering cross sections on nuclear targets.  

The nucleon density in the light (A$ \leq 20$) nuclei can be described by a harmonic oscillator (HO) density distribution \cite{Elt1,Oza}:
\begin{equation}
\label{3.3}
\rho_A (r) = \rho_1 \left(1+\frac{A/2-2}{3} \left( \frac{r}{\lambda}\right)^2 \right)
\exp\left(-\frac{r^2}{\lambda^2}\right) ,
\end{equation}
where $\lambda$ is the nucleus size parameter and $\rho_{1}$ is the normalization constant. The nuclear density distributions in not very light nuclei can be reasonably described by Woods-Saxon expression
\begin{equation}
\label{3.4}
\rho_A (r) = \frac{\rho_1}{1+\exp\left(\frac{r-c}{a}\right)}.
\end{equation}
Here $\rho_{1}$ is the normalization constant, $c$ is a parameter measuring the  nuclear size, and $a$ is related to the diffuseness of the nuclear surface. The parameter $c$ shows the value of $r$  at which $\rho(r)$ decreases by a factor 2 compared to $\rho(r=0)$,  $\rho(r=c) = \frac{1}{2}\rho (r = 0)$. The value of $a$ determines the distance $r = 4a \ln 3 \sim 4.4 a$ at which $\rho(r)$ decreases from $0.9\rho(r = 0)$ to $0.1\rho(r = 0)$.  

It was mentioned above that from the experimental data on $\sigma^{(I)}_{AB}$ it is possible to determine only one parameter of nuclear matter distribution, let us say ${\rm R_m}$, Eq.~(\ref{1.7}). It is necessary to note that the value of ${\rm R_m}$ is somewhat smaller than the effective radius $R_A \simeq 1.2 A^{1/3}$ fm.  For example, in the case of uniform nuclear density with radius $R_A$,  $R_A = \sqrt{\frac{5}{3}\langle r^2_A \rangle}$.  

The reaction cross section of  $^{12}$C-$^{12}$C interaction as a function of the ${\rm R_m}$ of the $^{12}$C nucleus was calculated in different approximations of the Glauber Theory for various nucleon density distributions. The results of the calculations are shown in Figure 3. 

\begin{figure}[th]
\includegraphics[width=0.7\textwidth]{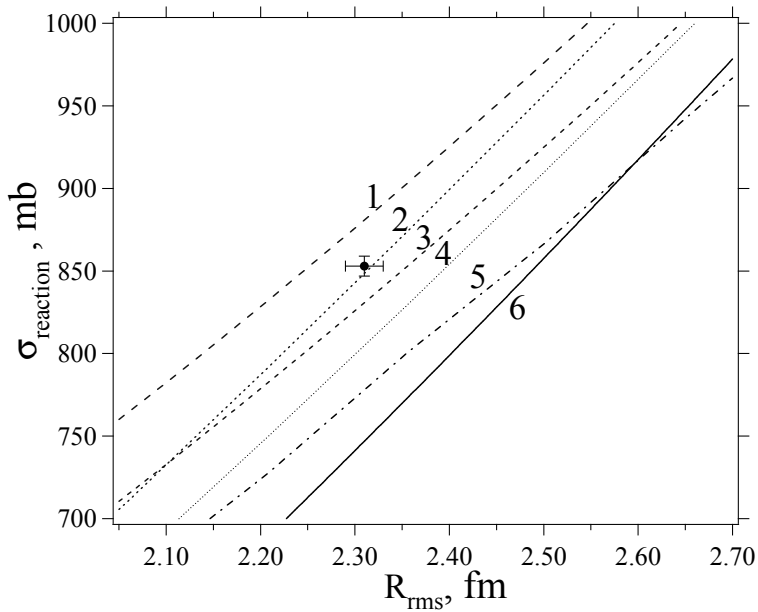}
\caption{The cross section of the reaction $^{12}$C$-^{12}$C as a function of the ${\rm R_{rms}}$ of the $^{12}$C nucleus calculated in different approximations with two different nucleon densities: 1) the optical approximation without range of NN interaction with Woods-Saxon density distribution; 2) the optical approximation without range of NN interaction with HO-potential density distribution; 3) the rigid target approximation without range of NN interaction with Woods-Saxon density distribution; 4) the rigid target approximation without range of NN interaction with HO-potential density distribution; 5) Glauber calculation with Woods-Saxon density distribution; 6) Glauber calculation with HO-potential density distribution.}
\end{figure}

In these calculations, which correspond to energies of 800-1000 MeV per  projectile nucleon, the total NN cross section value, averaged over $pp$  and $pn$ interactions, $\sigma^{tot}_{NN} =$ 43 mb was used.  The nuclear  matter distribution parameters $\lambda$, for harmonic oscillator (HO)  density in Eq.~(\ref{3.3}), and $c$, for the Woods-Saxon density in Eq.~(\ref{3.4}),  have been fitted to obtain the required ${\rm R_m}$, whereas the parameter  $a$ in Eq.~(\ref{3.4}) has been fixed to the value $a = 0.54$ fm.  

As the first step we have reproduced the result from~\cite{Oza} for  $\sigma^{(r)}_{^{12}C-^{12}C}$ in the optical approximation without NN  interaction range, using the harmonic oscillator (HO) density. Dependence  of the  reaction cross section, $\sigma^{(r)}_{^{12}C-^{12}C}$, on the  ${\rm R_m}$ is shown by curve 2 of Fig.~3. It is in a good agreement with the result from~\cite{Oza}, which is shown by the marker.  

Curve 1 in Fig.~3 represents the calculations of the dependence of  the reaction cross section $\sigma^{(r)}_{^{12}C-^{12}C}$ on ${\rm R_m}$,  which was done in the optical approximation with the Woods-Saxon density  distribution.  By comparing curves 1 and 2 of Fig.~3 one can see  that equal ${\rm R_m}$ with different assumptions about the nuclear  density distribution result in different reaction cross sections. In other  words this equivalently means that the same experimental reaction (or  interaction) cross section with different assumptions about nuclear density  distribution result in different ${\rm R_m}$, e.g. the assumption of  the Woods-Saxon density distribution in $^{12}$C nucleus leads to a smaller  value of ${\rm R_m}$ than the one obtained with harmonic oscillator density  distribution.
 
The corresponding results obtained in the rigid target approximation are shown by curves 4 and 3 in Fig.~3. Here again the assumption of Woods-Saxon density distribution results in a smaller value of ${\rm R_m}$ than the one calculated with the HO density distribution. The rigid target approximation contains additional diagrams Figs.~1d, 1e, 1f, 1g,..., which increase the shadow effects. That explains why both curves 3 and 4 lie below curves 1 and 2. 

The curves 5 and 6 in Fig.~3 show the results of the calculation of  all the diagrams of the Glauber Theory by using Monte Carlo method and accounting  for the finite range of NN interaction. Curves 5 and 6 were calculated with the Woods-Saxon density distribution and the HO density respectively. Here, new  shadow corrections of the type shown in Figs.~1h, 1i, 1j, 1k,..., appear by  comparison to the rigid target approximation. As a result, the calculated reaction cross section becomes now smaller at the same value of ${\rm R_m}$. 

Clearly, the optical-limit approximation overestimates the calculated nucleus-nucleus reaction cross sections. The difference between the reaction cross section calculated in the optical-limit approximation and that calculated with the help of the exact Glauber formula becomes even larger in the case of halo nuclei \cite{AlLo}. Therefore, to extract more accurate information on the nuclear sizes from the reaction cross sections, it is important to to perform calculations with the exact Glauber formula. It should be admitted however that the calculations of the nucleus-nucleus reaction cross sections by the exact Glauber formula require significantly more time compare to calculations in optical-limit approximation. The reaction cross sections calculations in the rigid-target approximation, being very simple, give noticeably more accurate results compare to the optical-limit approximation.

The values of the interaction cross sections $\sigma^{(I)}_{^{12}C-^{12}C}$  presented in~\cite{Oza} are $856 \pm 9$ mb and $853 \pm 6$ mb at energies 790 MeV and 950 MeV per nucleon, respectively. The older experimental measurement  in~\cite{Jar} gives a value $\sigma^{(I)}_{^{12}C-^{12}C} = 939 \pm 49$ mb at  energy 870 MeV per nucleon. The total  $^{12}$C-$^{12}$C cross section was measured to be $1254 \pm 54$ mb at the  same energy, whereas the Glauber Theory with ${\rm R_m}$ taken from data in~\cite{Oza} predicts a value $\sigma^{tot}_{^{12}C-^{12}C}$ = 1405 mb.

In Table~\ref{tab:table1}  we present the values of ${\rm R_m}$ extracted from the  measurements \cite{Oza1,Oza} of interaction cross section\footnote{Interaction cross section is defined \cite{Tani} as the total cross  section for the processes of nucleon (proton and/or neutron) $\sigma^{(I)}_{AB}$  removal from the  incident nucleus. The difference between $\sigma^{(I)}_{AB}$ and  $\sigma^{(r)}_{AB}$ was estimated \cite{KSh,Oga} to be less than a few percents  for a beam energy higher than several hundred MeV per nucleon and this difference  will be considered in details in Section~\ref{sec:section4}.} of stable projectile nuclei with  $^{12}C$ target at energies 800-1000 MeV per nucleon. 

The ${\rm R_m}$ values were calculated assuming that the nuclear matter  density distribution can be described by the Woods-Saxon expression Eq.~(\ref{3.4}).  To get the dependence of the reaction cross section $\sigma^{(r)}_{AB}$ on the  ${\rm R_m}$, we varied parameter $c$ of the density distribution and kept  parameter $a$ as a constant at $a = 0.54$ fm. The ${\rm R_m}$ values were  extracted from the agreement of the calculated value $\sigma^{(r)}$ with the  experimental values of $\sigma^{(I)}_{AB}$.

In Table~\ref{tab:table1} one can see that our calculations in the optical approximation  with the Woods-Saxon density distribution and neglecting the NN interaction  range result in the slightly smaller values of ${\rm R_m}$ (0.05$-$0.1 fm)  than those obtained in~\cite{Oza,Oza1}. ${\rm R_m}$ values are getting  even smaller when calculated with a finite range of NN interaction. In the  case of the Glauber Theory with the Woods-Saxon density distribution  shadow corrections lead to larger values of ${\rm R_m}$ than in the  other calculations. 

\begin{table}[pt]
\caption{\label{tab:table1}The values of ${\rm R_m}$ in fm extracted from the measurements of interaction cross section in collisions of projectile nuclear beam with $^{12}$C target at energies 800-1000 MeV per nucleon. Data on measured interaction cross section were taken from \cite{Oza1,Oza}.}
{\begin{tabular}{@{}cccccc@{}} \toprule
Nucleus & \multicolumn{2}{c}{Without NN range} & 
\multicolumn{2}{c}{With NN range} & Glauber Theory\\
& OH, \cite{Oza1,Oza} & Optical & Optical & Rigid target &  \\  \colrule
C$^{12}$ & $2.31 \pm 0.02$ & $2.25 \pm 0.01$ & $2.09 \pm 0.01$ & $2.18 \pm 0.01$ & $2.49 \pm 0.01$ \\ 
N$^{14}$ & $2.47 \pm 0.03$ & $2.42 \pm 0.03$ & $2.23 \pm 0.03$ & $2.35 \pm 0.04$ & $2.64 \pm 0.03$ \\ 
O$^{16}$ & $2.54 \pm 0.02$ & $2.48 \pm 0.02$ & $2.29 \pm 0.02$ & $2.41 \pm 0.03$ & $2.69 \pm 0.02$ \\ 
F$^{19}$ & $2.61 \pm 0.07$  & $2.55 \pm 0.08$ & $2.34 \pm 0.08$ & $2.44 \pm 0.09$ & $2.75 \pm 0.07$ \\ 
Ne$^{20}$ & $2.87 \pm 0.03$ & $2.84 \pm 0.04$ & $2.63 \pm 0.03$ & $2.75 \pm 0.04$ & $2.99 \pm 0.03$ \\ 
Na$^{23}$ & $2.83 \pm 0.03$ & $2.73 \pm 0.04$ & $2.52 \pm 0.04$ & $2.62 \pm 0.04$ & $2.91 \pm 0.03$ \\ 
Mg$^{24}$ & $2.79 \pm 0.15$ & $2.65 \pm 0.23$ & $2.44 \pm 0.22$ & $2.53 \pm 0.24$ & $2.85 \pm 0.20$ \\ 
Cl$^{35}$ & $3.045 \pm 0.037$ & $2.92 \pm 0.04$ & $2.68 \pm 0.04$ & $2.76 \pm 0.04$ & $3.08 \pm 0.04$ \\ 
Ar$^{40}$ & $3.282 \pm 0.036$ & $3.16 \pm 0.04$ & $2.90 \pm 0.03$ & $2.98 \pm 0.04$ & $3.30 \pm 0.03$ \\ \botrule
\end{tabular}}
\end{table}

\subsection{\label{sec:section3.3}Comparison of the calculated radii of matter distribution, ${\rm R_m}$, and radii of charge distribution, ${\rm R_{ch}}$}

Let us compare the values for ${\rm R_ m}$ obtained from  nucleus-nucleus collisions and presented in Table~\ref{tab:table1} with the published results. It is known  from \cite{CLS,ABV} that radii of proton and neutron distributions in  nuclei with $Z \simeq A/2$ are practically equal, so we can compare  calculated radii for nuclear matter, ${\rm R_m}$, with electrical charge radii ${\rm R_{ch}}$ presented in~\cite{DeDe}.

It is necessary to make a distinction between the distributions of the centers of nucleons $\rho_A(r)$ and the folded distributions $\tilde {\rho}_A(r)$, where density $\rho_A(r)$ is convoluted with the matter or charge density in the nucleon, $\rho_N(r)$:
\begin{equation}
\label{eq20}
\tilde{\rho}_A(r) = \int  \rho_A(r-r_1) \rho_N(r_1) d^3r_1 \;.
\end{equation}

\noindent
We have to deal with $\rho_A(r)$ and with $\tilde{\rho}_A(r)$ when we calculate  ${\rm R_m}$ with and without accounting for the range of NN interaction, respectively. The rms radii of $\tilde{\rho}_A(r)$ and $\rho_A(r)$, $\tilde{\rm R}_m$ and ${\rm R_m}$, are different and the following relation between them was used in~\cite{CLS}:
\begin{equation}
\label{eq21}
\tilde{\rm R}_m^2 = {\rm R_m^2} +(0.82 \hspace{0.1cm}{\rm fm})^2 \;.
\end{equation}

In Fig.~4 we compare the result of ${\rm R_m}$ calculations in the optical approximation (triangles) and in the rigid target approximation (squares) with $\tilde{\rm R}_{ch}$ values extracted from the electron-nucleus scattering experiment,~\cite{DeDe}. Calculations of ${\rm R_m}$  in both approximations  were done with zero range of NN interaction (i.e. the folded distribution $\tilde{\rho}_A(r)$ was used) and Woods-Saxon density distribution. Obtained  results are systematically smaller than the data presented in~\cite{DeDe}.  This supports our point of view that in the optical approximation and in the  rigid target approximation the effects of nuclear shadowing are too small, and,  thus, one obtains agreement to the experimental nucleus-nucleus cross section  with a smaller value of $\tilde{\rm R}_m$. 

\begin{figure}[th]
\includegraphics[width=0.7\textwidth]{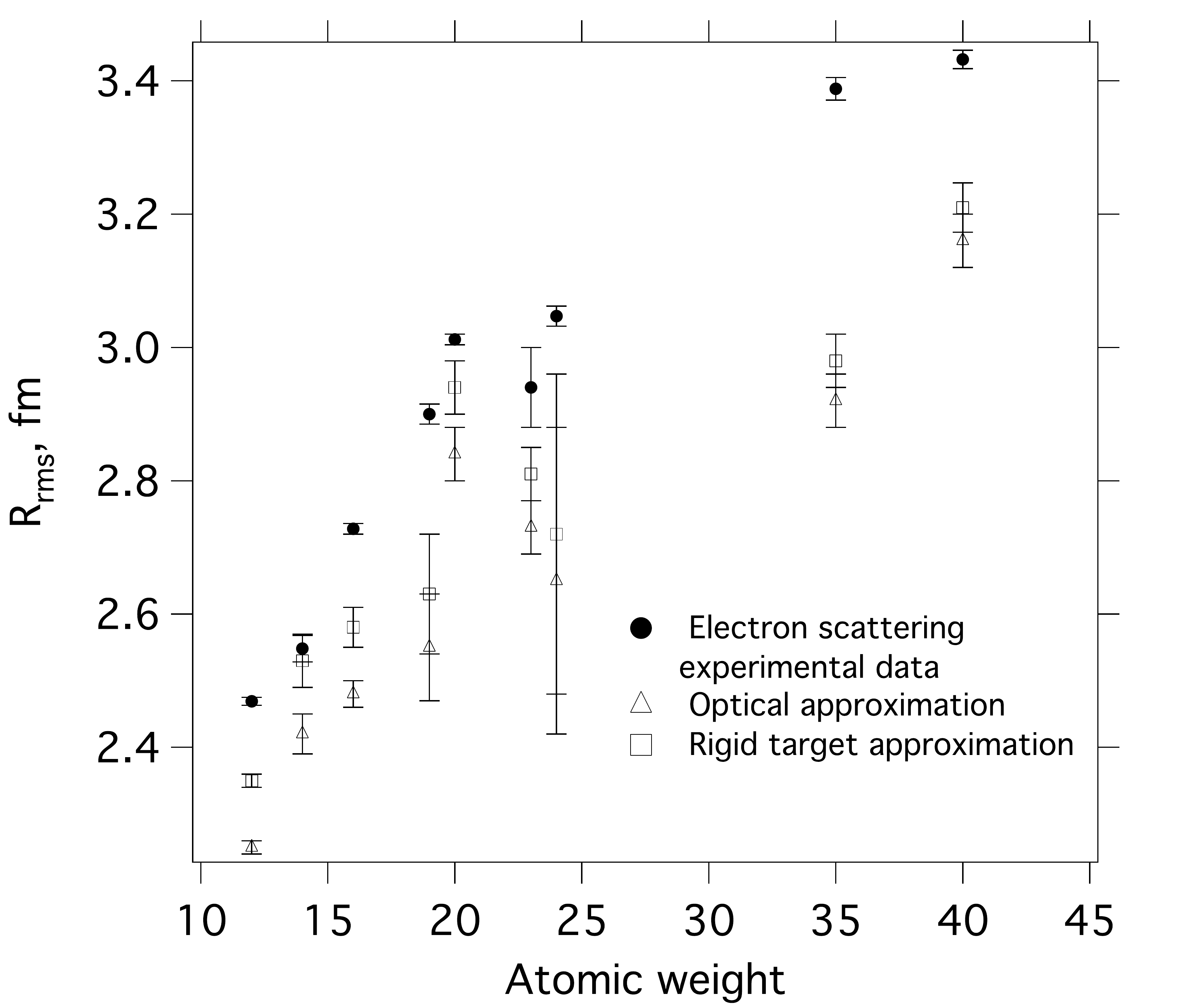}
\caption{The values of ${\rm R_{ch}}$ extracted from electron-nucleus scattering (filled circles) and the values of ${\rm R_m}$ obtained from nucleus-nucleus collisions in the optical approximation (triangles), and in the rigid target approximation (squares), both with zero range of NN interaction and with Woods-Saxon density distribution.}
\end{figure}

The same calculations of ${\rm R_m}$ were done in the framework of the Glauber Theory. The results of these calculations are presented in Fig.~5. In this  case it is impossible to provide calculations with zero range of NN  interaction because the contributions of diagrams with loops (see for  example, Fig.~1k) present this range in the denominator.

The ${\rm R_m}$ values calculated with distribution $\rho_A(r)$ together  with the $\tilde{\rm R}_{\rm ch}$ values extracted from electron-nucleus  scattering experiment are presented in Fig.~5. They are in slightly better  agreement to electron-nucleus scattering data than in the case shown in Fig.~4. 

\begin{figure}[th]
\includegraphics[width=0.45\textwidth]{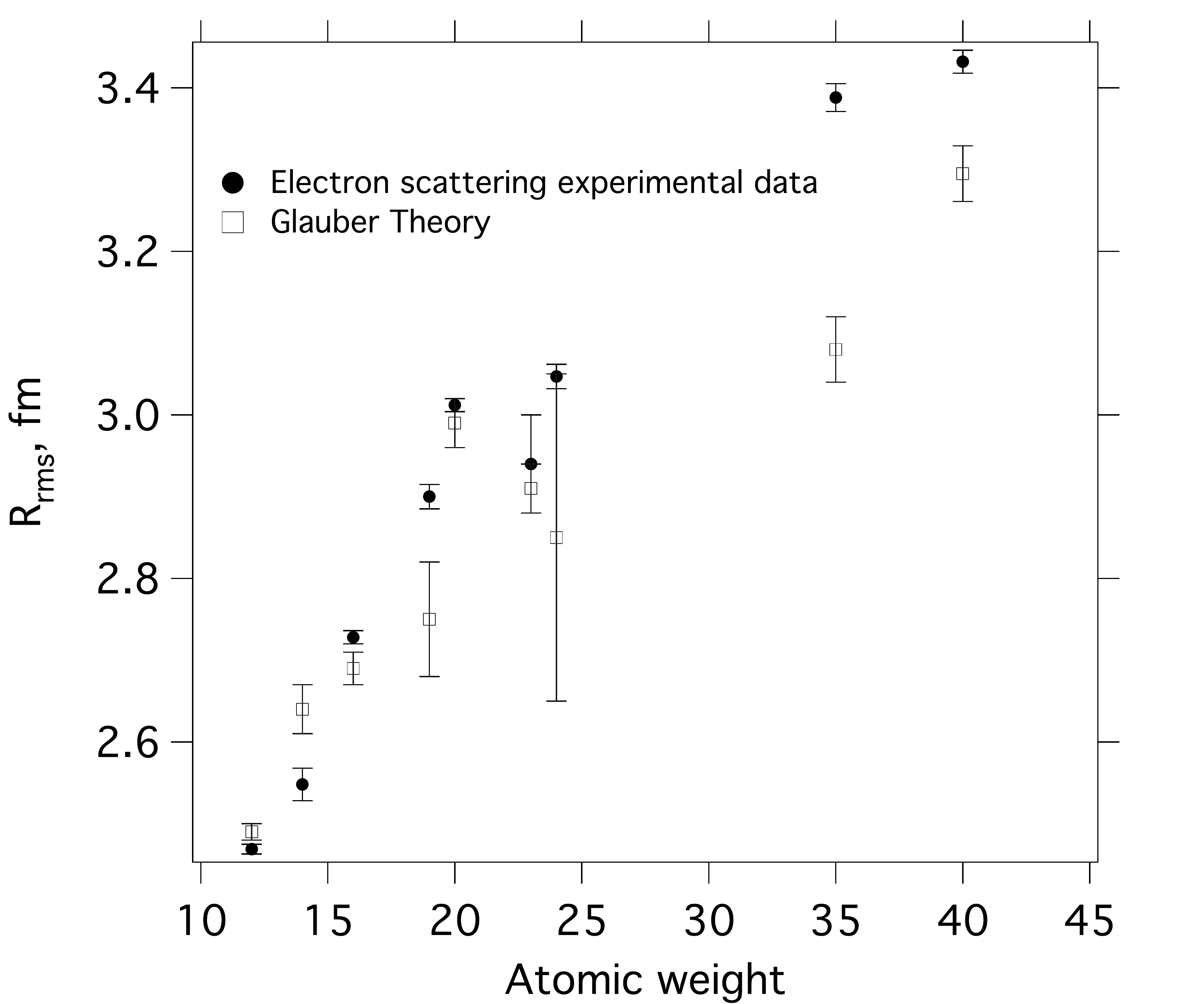}
\includegraphics[width=0.45\textwidth]{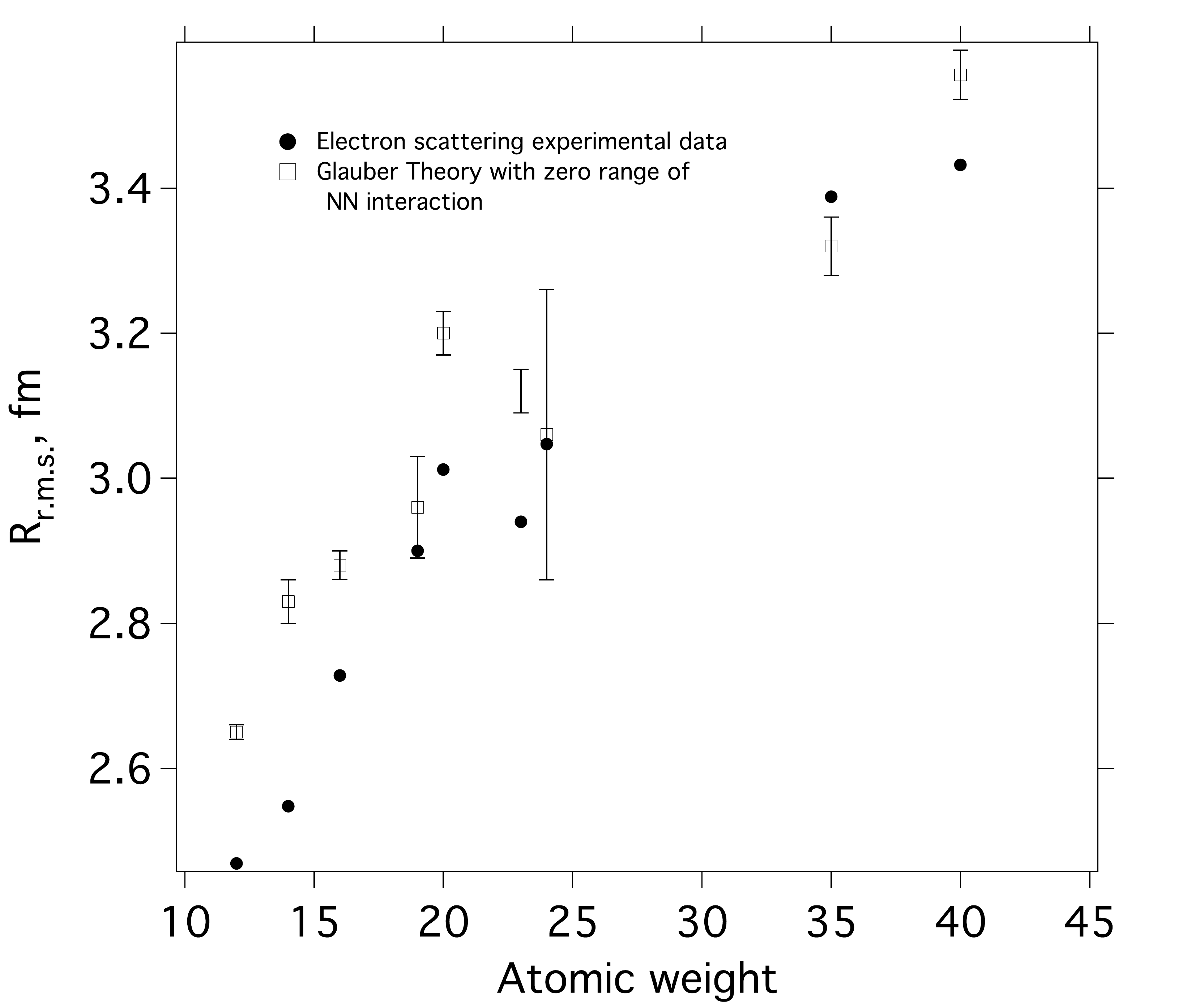}
\caption{The values of ${\rm R_{ch}}$ extracted from electron-nucleus scattering (filled circles) and the values of ${\rm R_m}$ obtained from nucleus-nucleus collisions in the Glauber Theory (open square) with Woods-Saxon density distribution and with finite range of NN interaction.}
\end{figure}

However, as we discussed before, the electron scattering data are related to the folded distributions  $\tilde {\rho}_A(r)$ \footnote{We neglect the difference of electromagnetic and strong interaction nucleon radii.}. In order to make a more reasonable comparison, we calculate the needed  ${\rm \tilde{R}_m}$ in the case of the Glauber Theory as  ${\rm R_m} + \Delta$,  where $\Delta$ was calculated as a difference  between the ${\rm R_m}$ obtained with the distribution of the nucleon  centers and ${\rm R_m}$ obtained with the nuclear matter density  distribution, Eq.~(\ref{3.4}). Both root-mean-square radii were calculated in  the optical approximation. 

\renewcommand{\theequation}{\arabic{section}.\arabic{equation}}
\setcounter{equation}{0}
\section{\label{sec:section4}Unstable nuclei, halo and skin}

\subsection{\label{sec:section4.1}Discovery of neutron halo}

The new era in nuclear physics was started in 1985, \cite{Tani,Tani1}, when the unstable nuclei with very large interaction cross sections were discovered. As the example of these sensation results we present in Table~\ref{tab:table2} the part of tables taken from \cite{Tani,Tani1} on the interaction cross section of helium and lithium isotopes.

\begin{table}[pt]
\caption{\label{tab:table2}Interaction cross sections $\sigma_I$ of projectile nucleus on $^{12}$C target at 790 MeV/nucleon.}
{\begin{tabular}{@{}cc@{}} \toprule
Nucleus & $\sigma_I$ (mb) \\ \colrule
$^3$He & $550 \pm 5$  \\   
$^4$He & $503 \pm 5$  \\   
$^6$He & $722 \pm 6$  \\   
$^8$He & $817 \pm 6$  \\   
$^6$Li & $688 \pm 10$  \\   
$^7$Li & $736 \pm 6$  \\   
$^8$Li & $768 \pm 9$  \\   
$^9$Li & $796 \pm 6$  \\  
$^{11}$Li & $1040 \pm 60$  \\   \botrule
\end{tabular}}
\end{table}

It is evident that when the interaction cross sections for lithium isotopes with atomic mass in the range from 6 to 9 show some regular behavior, the cross section of $^{11}$Li shows somewhat significant increase in its magnitude. Similar behavior was observed for helium isotopes.  The simplest expression for the interaction cross section Eq.~(\ref{1.4}) together with the assumption that  all these nuclei are spherical was used to analyze experimental data, \cite{Tani4}. The results taken from \cite{Tani1} for the radii of helium and lithium isotopes are presented in Fig.~6.  The significant increase of $^{11}$Li radius in comparison with nearest nuclei is evident and remains valid even after accounting for the criticism presented in \cite{ElTa}. The classical behavior of nuclear radii A-dependences shown in Eq.~(1.2) is evidently violated. 

\begin{figure}[th]
\includegraphics[width=0.6\textwidth]{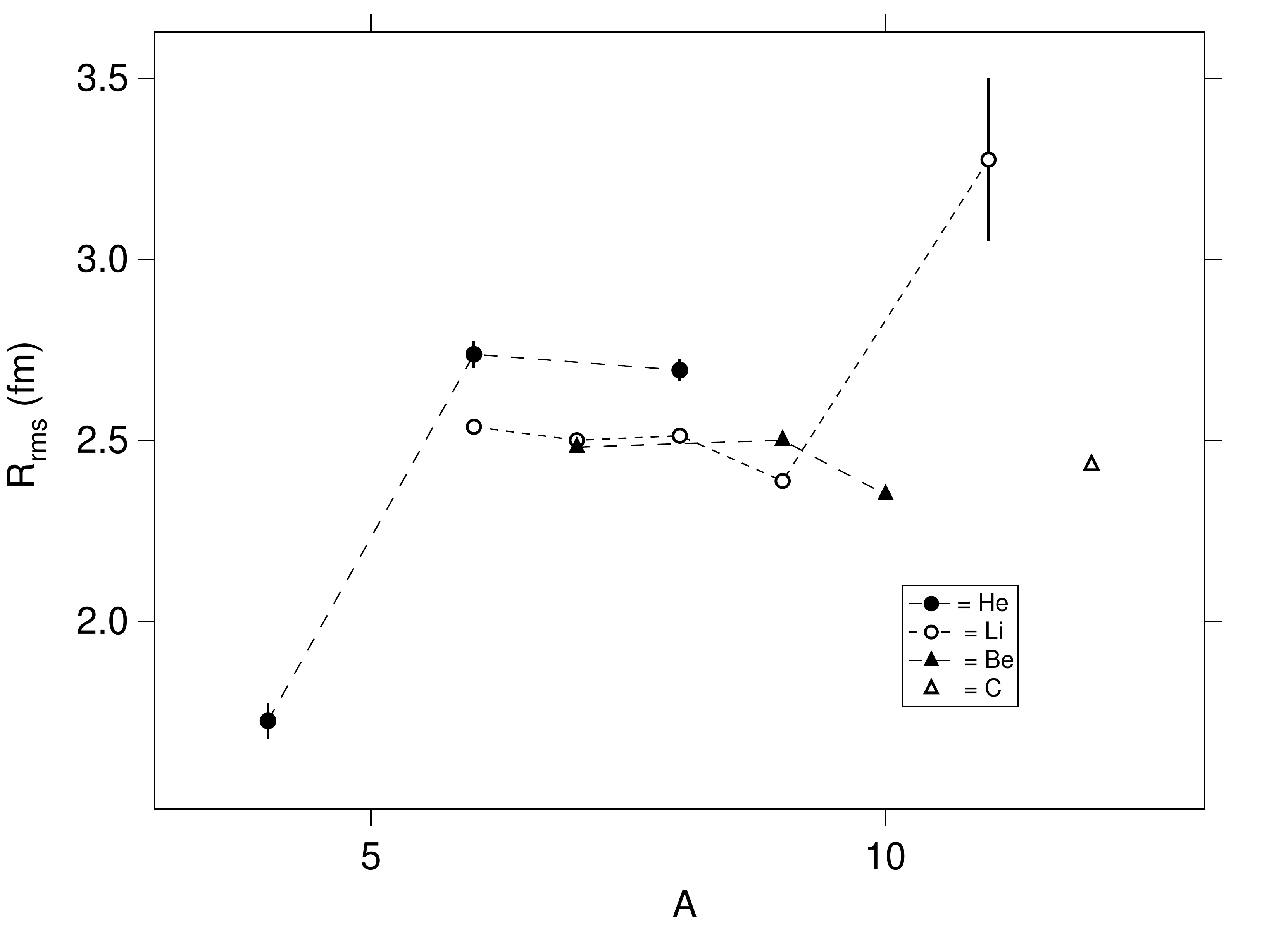}
\caption{ Matter rms radii ${\rm R^{(m)}_m}$. Lines connected isotopes are only guides for the eye. Differences in radii are seen for isobars with A = 6, 8 and 9. The $^{11}$Li isotope has a much larger radius than other nuclei.}
\end{figure}

The Hartree-Fock varionatial calculations with Skyrme potential for the  structures of light nuclides are presented in \cite{Sato}. It was shown that calculated interaction cross sections reproduce experimentally measured ones for all considered light nuclei except of $^{11}$Li, where the calculated cross section is considerably smaller than the experimental one\footnote{Later the interaction cross  sections of neutron-rich nuclei were described by Glauber Theory calculations  with nuclear densities obtained from Skyrme Hartree-Fock calculations with  accounting for more detailed picture, see, for example, \cite{BTR}.}.

The experimental results obtained in \cite{Tani,Tani1} and surprisingly large value for the $^{11}$Li radius were interpreted in \cite{HanJon} as an evidence of a neutron halo existence appearing as a result of very low neutron binding energy of the valence neutrons.

The nucleus $^{11}$Li can be considered as a system of $^9$Li core and two neutrons located in the halo -- long tail of nuclear density distribution. In \cite{Tani6} it was assumed that the density of $^{11}$Li nucleus can be written as
\begin{equation}
\label{4.1}
\rho_{^{11}Li}(r) = \rho_c(r) + \rho_h(r) \;,
\end{equation}
where indices $c$ and $h$ are related to core and halo contributions. Using this assumption following values for $\rm R_m$ values were obtained: 
$\rm R_m (^{11}Li) = 3.1 \pm 0.3$ fm, $\rm R_m(c) = 2.5 \pm 0.1$ fm and $\rm R_m(h) = 4.8 \pm 0.8$ fm. 
 
$^6$He nucleus also can be considered as $^4$He core and two neutrons located in the halo. Evidently, every neutron has a probability to belong to a core or a halo. 

Significant contribution of three-body forces between the core and two neutrons is a possible explanation of the anomaly large r.m.s. radius of $^{11}$Li. An assumption that the $^{11}$Li ground state should be considered as a three-body $^9$Li+n+n system was confirmed later in many papers, see, for example, the analyses in \cite{EHMS,Shul}. An estimation of $^{11}$Li wave function is presented in \cite{Shul}. 

It was pointed out (see, for example, in \cite{FJR}) that three particles interacting via short-range two-body interaction can form a variety of structures. The qualitative picture for the case of (A+n+n) system (core nuclide A and two neutrons) is shown in Fig.~7 (picture is taken from \cite{HJJ}). In such systems a so-called Borromean states can exist. Borromean state can be defined as a bound three-body system in which none of the two-body subsystems form a  bound state. It is necessary to distinguish \cite{FJR1} Borromean states from Efimov's systems \cite{Efi}.

\begin{figure}[th]
\includegraphics[width=0.6\textwidth]{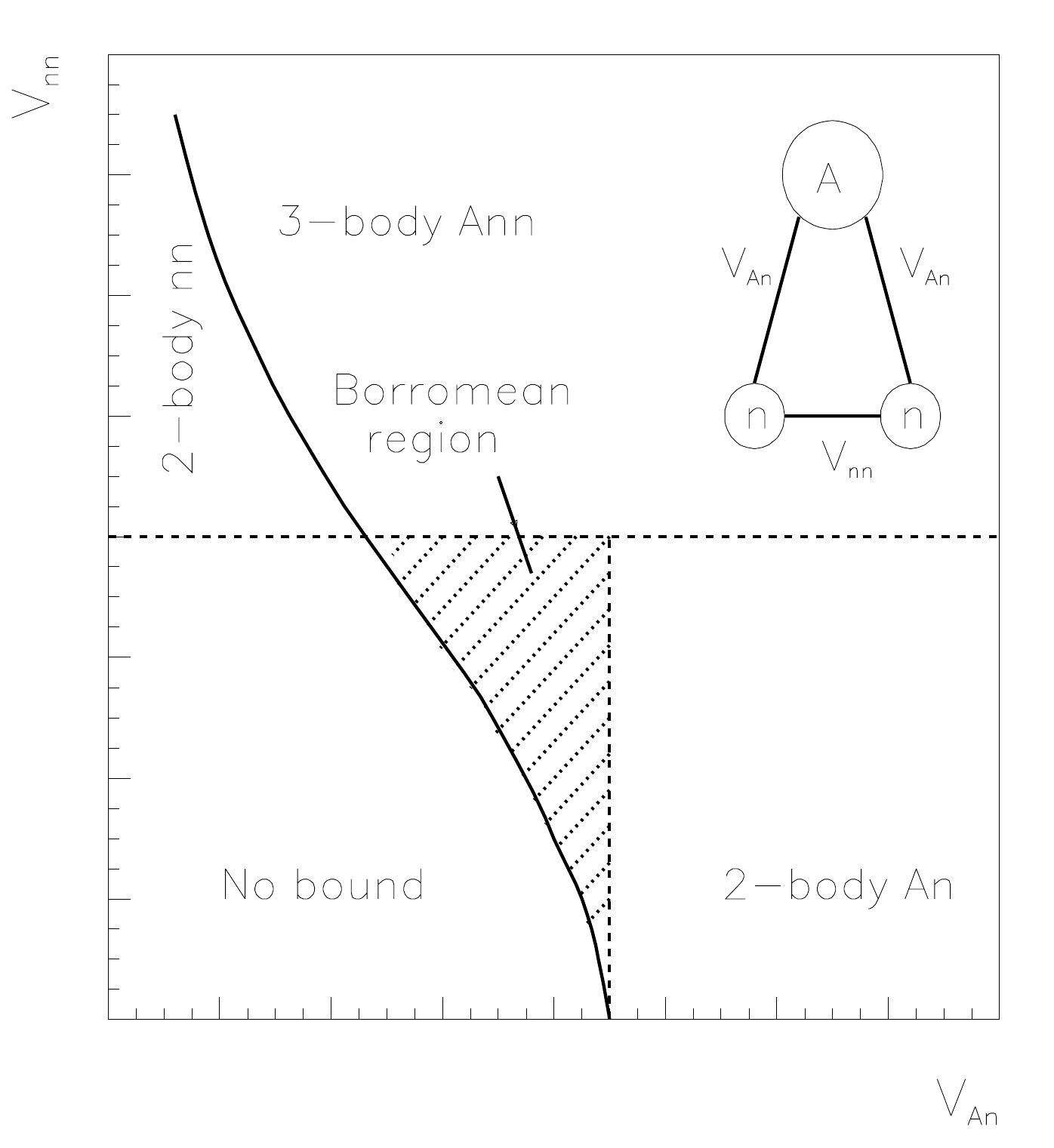}
\caption{Schematic classification of three-body states of the system (A+n+n) as a function of the strengths of the two-body potentials $V_{An}$ and $V_{nn}$. The curve separates the regions where the three-body system is either bound or unbound, and the dashed lines separate bound and unbound two-body systems. The Borromean region is shown as dashed area.}
\end{figure}

Both $^{6}$He and $^{11}$Li nuclei are considered as the excellent  examples of a Borromean system confirmed experimentally. Both nuclei are stable in relation to the strong interaction, whereas nuclides $^{5}$He and $^{10}$Li as well as the dineutron {\it nn} state are the unbound systems. 

It is interesting to note that stable nuclei can have the Borromean excited states. As an example \cite{YFH}, let us consider the Hoyle resonance in $^{12}$C ($0^+$ state with energy 7.65 MeV). $^{12}$C nucleus can be considered a cluster of three $\alpha$-particles and every two-body system here is unbound $^8$Be state. This Hoyle resonance plays very important role in  stellar nucleosynthesis \cite{Hoyle,Hoyle1,Livio} and, in some sense,  the existence of our Universe is based on its existence.

In addition to the Borromean states, the three-body system can exist in the configurations, first one where one two-body system is bound and two are unbound ("Tango" configuration, \cite{Rob}) and another where two two-body system are bound and one is unbound ("Samba" configuration, \cite{YTF}). It is possible, that $^{20}$C nuclide is an example of "Samba" system, which is composed of two-neutron halo and a $^{18}$C core. The neutron and $^{18}$C forms the weakly bound state $^{19}$C. 

\subsection{\label{sec:section4.2}Neutron (proton) halo and skin}

After the discovery of several unstable nuclei with unusual nuclear matter distribution, the existence of neutron or proton halo \cite{Tani,Tani1} and neutron or proton skin \cite{Tani12} became clear.

Usually, halo is considered as a long low density tail in the nuclear matter distribution \cite{Riis}, whereas skin means a significant difference  between R$_{\rm m}$ values for protons and neutrons. The difference between these two scenarios depends on the values of some parameters which can be  determined in theoretical Hartree-Fock calculations. Sometimes same nuclei are considered by different authors to have halo or skin. For example, in \cite{Tani12} authors described nuclides $^6$He and $^8$He as ones with nuclear skin, meanwhile in \cite{Ryj} same nuclei are described as nucleon with two-nucleons and four-nucleons halos. Very clear outlook of this problem was presented in \cite{Tani5,Tani9}. 

Firstly, let us consider a nucleus with given values of Z and N which nuclear density distribution exhibits the presence of neutron skin. Usually the ratio of neutron to proton density  distributions, $\rho_n(r)/\rho_p(r)$, is about N/Z in the interior region of a nucleus. This ratio can somewhat exceed N/Z near the nuclear surface, which can be explained by fluctuations due to shell effects, etc. The following criteria were introduced in \cite{Tani5} to define the neutron skin:
\begin{enumerate}
\item  in the neutron skin
\begin{equation}
\label{4.2}
\rho_n(r)/\rho_p(r) > \xi_1 \;\; \rm for \;\; r \sim R_A \;.
\end{equation}
It was recommended in \cite{Tani5} to use $\xi_1$ = 4.
\item contrary to the case of neutron halo, a neutron skin should contain a significant number of neutrons.
\begin{equation}
\label{4.3}
\rho_n(r \sim R_A)/\rho_n(r=0)  > \xi_2  \;.
\end{equation}
Again, it was suggested in \cite{Tani5} that the value of $\xi_2$ should be 1/100
\item the difference between neutron and proton radii should be large enough, i.e.
\begin{equation}
\label{4.4}
\delta R  = R_n - R_p > \xi_3  \;,
\end{equation}
where $\xi_3 = 1$ fm \cite{Tani5}.
\end{enumerate}

The existence of nuclei with proton skin is also under discussions. The proton skin can be defined in the same manner as it was done for the neutron skin in Eqs.(\ref{4.2})-(\ref{4.4}). $^8$B and $^{21}$Al are amongst the possible nuclei with proton halo. In the case of $^{21}$Al the value $\delta R$ defined as in Eq.~(\ref{4.4}) is equal to -0.84 fm \cite{Tani5}. Generally, proton halos are equally possible as a neutron halos but less pronounced because of the strong influence of the Coulomb barrier.

Definitions of halo and skin presented in \cite{Tani5} suggest that the main difference between nuclear skin and nuclear halo is the second condition, see  Eq.~(4.3). The skin is a phenomenon which involves rather large number of nucleons, whereas halo is caused by one or two nucleons extremely loosely bound with the core, that results in an abnormal slope of nuclear density distribution tail.  

The general outlook on light nuclei with halo or skin is presented in Fig.~8 taken from \cite{Khal}.

\begin{figure}[th]
\includegraphics[width=0.6\textwidth]{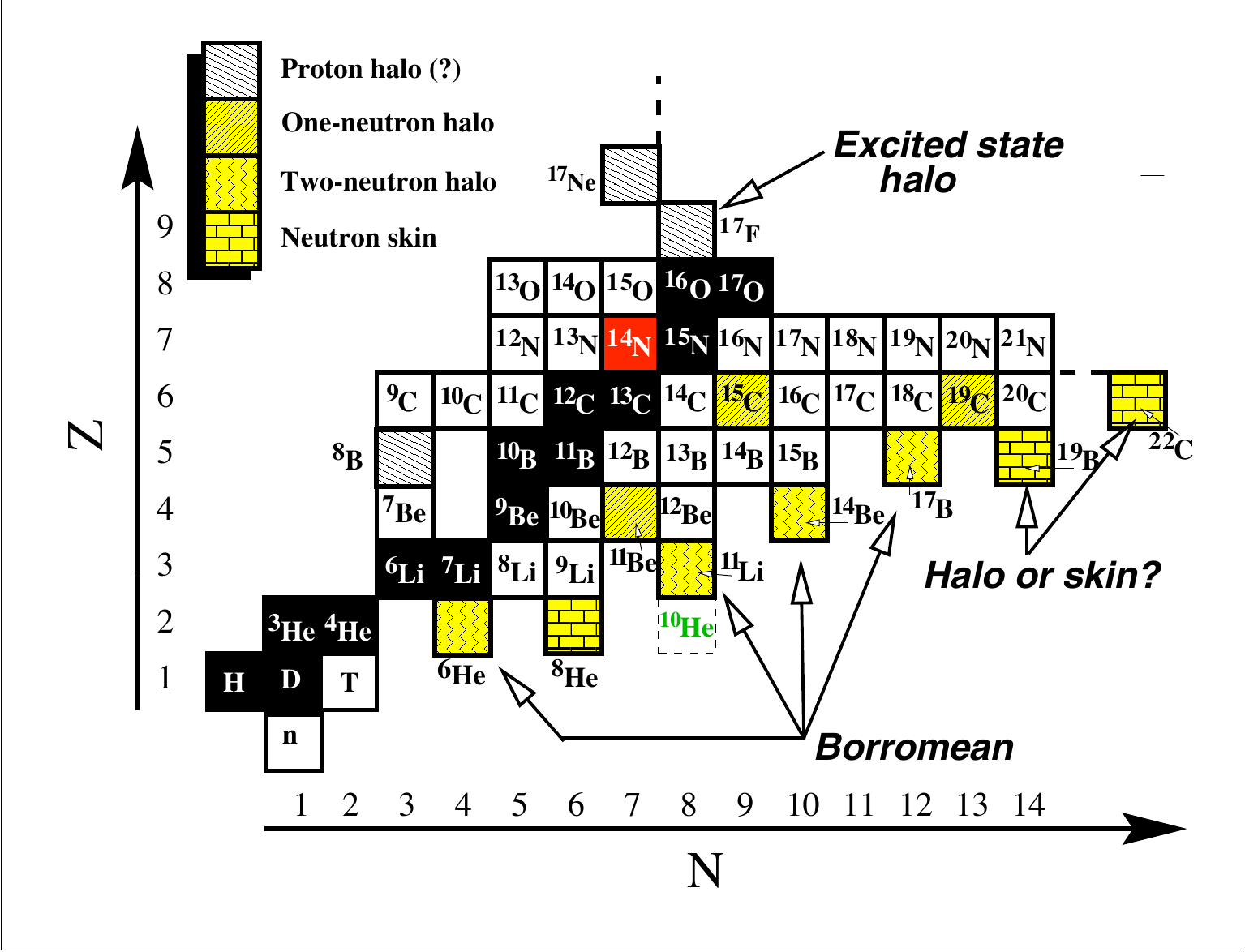}
\caption{The light end of the chart of nuclides. Some of the drip line nuclides found to exhibit new phenomena, such as halos and skin. }
\end{figure}

It seems that the skin/halo effects decrease with increase of atomic weight. For example, the cross section or interaction radii measured in the  collisions of Na isotopes on a carbon target \cite{Tani10} in a wide range of atomic weights (A=20-23, 25-32) increase with the number of neutrons  rather regularly, contrary to the behavior of He and Li isotopes shown in  Fig.~6. The neutron skin in neutron-rich Na isotopes is only slightly larger in comparison with skin in Ca isotopes, \cite{ABV}.

The theoretical analysis provided in \cite{HJJ} results in the necessary, but not sufficient condition for occurrence of two- or three-body halo systems (i.e., say, one-neutron or two-neutron halos):
\begin{equation}
\label{4.5}
S_N A^{2/3} \leq (2 \div 4) \; {\rm MeV} \; ,
\end{equation}
where $S_N$ is one-nucleon or two-nucleon separation energy and A is atomic weight.

Detailed discussion of neutron halo nuclei can be found in \cite{Tani7}. The theory of nuclear halos can be found in \cite{Zhu,Bert}.

The so-called few-body approach of the Glauber model was suggested in \cite{ATT,TA,AOST} for the analyses of reactions for halo nuclei.  The nucleus with halo, say  $^6$He or $^{11}$Li, was considered as the systems of core (c), $^4$He or $^9$Li, and two valence ($v$) neutrons. The nuclear density distribution  of the halo projectile (P) nuclide can be written as 
\begin{equation}
\label{4.6}
\rho_P(r) = \rho_c(r) + \rho_v(r)  \;,
\end{equation}
where the relative motion of core and valence neutrons is accounted for.  If the interactions of core and valence nucleons are calculated in the  optical limit of the Glauber Theory, the few-body approach theoretical  expressions are different, \cite{TJA}, from the standard optical limit,  so the numerical results are also different. However, in the complete Glauber  Theory the results of calculations in the few-body approach for $^6$He-$^{12}$C  and $^{11}$Li-$^{12}$C interactions practically coincide up to numerically  small contributions coming from nucleon correlations in the few-body approach \cite{AlLo} with the results of the standard Glauber expression. 

\subsection{\label{sec:section4.3}Nucleon separation energies and fragmentation cross sections for unstable nuclei}

Let us consider an unstable nucleus as a core and one nucleon distributed at somewhat long distance from the core (in units of the range of the strong interaction). The probability $\rho(r)$ to find such nucleon at distant $r$ from the centre of a system is defined in quantum mechanics \cite{LaLi}, \cite{LISchiff} by the  square modulus of the correspondent wave function 
\begin{equation}
\label{4.7}
\rho(r) \sim  \frac1r e^{-2kr} \;,
\end{equation}
where 
\begin{equation}
\label{4.8}
k =  \sqrt{2m \epsilon} \;.
\end{equation}
Here $\epsilon$ is the bound energy and $m$ is a nucleon mass. 

Evidently, the nuclides with halo should have a small separation energy of a  neutron or a proton, $S_n$ or $S_p$, in comparison with stable nuclei, where  this energy is about 6-8 MeV \cite{Tani8}. 

The nucleus of $^{11}$Li can be used as a clear illustration of this phenomenon. Usually $^{11}$Li is considered as a core $^9$Li with two  neutron halo (long distance tail in the $^{11}$Li wave function). The separation energy of two neutrons from $^{11}$Li nuclide defined as 
\begin{equation}
\label{4.9}
S_{2n}(^{11}\rm Li) = \Delta E[ (^9\rm Li + 2n) - (^{11}\rm Li)] 
\end{equation}
is extraordinary small
\begin{center}
$S_{2n}(^{11}\rm Li) = 0.34 \pm 0.04$ MeV \cite{Tani8} \\
$S_{2n}(^{11}\rm Li) = 0.295 \pm 0.035$ MeV \cite{You} \\
$S_{2n}(^{11} \rm Li) = 369.15 \pm 0.65$ keV \cite{Smi}. 
\end{center}

The main candidates for being nuclei with neutron or proton halo together with their one-neutron and one-proton separation energies, and  mean life times are presented in Table~\ref{tab:table3}. In some cases, for example, for $^6$He and $^{11}$Li nuclides, the one-neutron and two-neutron separation  energies coincide because the nuclides $^5$He and $^{10}$Li are unstable and the second neutron is emitted during a nuclear time-scale.

\begin{table}[pt]
\caption{\label{tab:table3}Possible halo states.}
{\begin{tabular}{ccccc} \toprule
Nuclide   & $S_n$ (keV)      & $S_p$ (keV) & $\tau$ & Configuration  \\ \colrule
$^6$He    &  $1867 \pm 50$  & $ 26520 \pm 95$  & $807 \pm 15$ ms    & $^4$He + 2n \\   
$^8$He    &  $2574 \pm 18$  & -                & $119 \pm 12$ ms    & $^4$He + 4n \\   
$^{11}$Li &  $325 \pm 31$   & $ 15303 \pm 31$  & $8.59 \pm 0.14$ ms & $^9$Li + 2n \\   
$^{11}$Be &  $504 \pm 6$    & $ 20165 \pm 16$  & $13.8 \pm 0.8$ s   & $^{10}$Be + n \\   
$^{14}$Be &  $1850 \pm 51$  & -                & $4.35 \pm 0.17$ ms & $^{12}$Be + 2n \\   
$^8$B     &  $13020 \pm 70$ &  $137 \pm 1$     & $770 \pm 3$ ms     & $^7$Be + p \\   
$^{17}$B  &  $1430 \pm 150$ & -                & $5.08 \pm 0.95$ ms & $^{15}$B + 2n \\   
$^{19}$C  &  -              & -                & $49 \pm 4$ ms      & $^{18}$C + n \\   
$^{17}$F  &  $16800 \pm 8$  & $ 600.3 \pm 0.2$ & $64.49 \pm 0.16$ s & $^{16}$O + p \\   \botrule
\end{tabular}}
\end{table}

Generally, the halo nuclei have one-nucleon or two-nucleons separation energies less or approximately 1 MeV. However, there are some exceptions. Amongst these exceptions are helium isotopes, which neutron separation energies are 1.9 MeV for $^6$He and 2.6 MeV for $^8$He. That permits us to say that these nuclides have neutron skins.

At the same time the nuclide $^{14}$B has one-neutron separation energy  $S_n = 0.97$ MeV, but it is not considered as a halo state. Possibly,  the reason is that in beta-decay of $^{14}$B the modes $\beta^-n$ and  $\beta^-nn$ are not observed, contrary to the cases of nuclides  presented in Table~\ref{tab:table3}, where such decays exist. Even the halo state of  $^8$B where the proton separation energies is smaller than 0.14 MeV is  not considered as evident, \cite{Riis}. All these facts indicate that today there is no the formal scheme which would allow to determine halo and skin nuclides. On the other hand, the two-neutron halo structure of $^6$He, $^{11}$Li, and $^{14}$Be is successfully reproduced by theoretical  calculations \cite{YTF}. 

Let us consider now the cross sections of fragmentation of the unstable nucleus with halo (A) into fragment F, which is the core of A, after collision with stable nucleus (B) from a target. The fragmentation process can be expressed as $AB \to F+X$, where X consists of all possible states of halo nucleons together with possible production of secondaries. All states of target nucleus are also included in X.

The  nuclear density of a halo nucleus can be considered similarly to Eq.~(4.6). Using expressions Eq.~(\ref{2.16})-(\ref{2.19}), the difference between reaction cross sections $\sigma^{(r)}_{AB}$ and $\sigma^{(r)}_{FB}$ can be written as
\begin{eqnarray}
\label{4.10}
\sigma^{(r)}_{AB} - \sigma^{(r)}_{FB} = \int d^2b \, && e^{-\sigma_{NN}^{tot} \int d^2b T_c(b-b_1) T_B(b_1)} \times \nonumber \\
&& \times \left[1 - e^{-\sigma_{NN}^{tot} \int d^2b T_h(b-b_1) T_B(b_1)}\right] \;.
\end{eqnarray}

The obtained expression can be viewed on as probability of the process in which the core of nucleus A doesn't interact and nucleons from halo does not interact. The probability of the former is expressed as exponential factor in (\ref{4.10}) and the probability of the latter is in square brackets.

So, it should correspond to the cross section of AB $\to$ F+X fragmentation cross section 
\begin{equation}
\label{4.11}
\sigma(AB \to F+X) = \sigma^{(r)}_{AB} - \sigma^{(r)}_{FB} \;,
\end{equation}
in accordance with the result presented in \cite{Oga} for the case when the  beam nucleus is $^{11}$Li and the fragment nucleus is $^9$Li. 

In Table~\ref{tab:table4} we compare the fragmentation cross sections with the differences of beam and fragment reaction cross sections. Since expression Eq.(\ref{4.11}) for $\sigma(AB \to F+X)$ was obtained in the optical approximation without accounting for the shadow effects between core and halo distribution, it is hard to expect perfect match with experimental data. Nevertheless, one can see a reasonable agreement of the experimental data with calculations done using expression Eq.~(\ref{4.11}) for the cases of $^{14}$Be $\to ^{12}$Be,  $^{11}$Be $\to ^{10}$Be, $^{11}$Li $\to ^9$Li and $^6$He $\to ^4$He fragmentation.  

The absence of agreement between experimental data and calculation done using Eq.~(\ref{4.11}) for $^8$He $\to ^4$He and  $^8$He $\to ^6$He fragmentation requires additional explanation.  After interaction of one or several valence neutrons of $^8$He with the target, two bound states, $^6$He and $^4$He, can appear after the final state interactions. So, one can assume that the sum of two and four  neutron separation cross sections of $^8$He nuclide should be very close to the difference in interaction cross sections of $^8$He and $^4$He: 
\begin{center}
\[ \sigma(^8{\rm He} \to ^6{\rm He}) + \sigma(^8{\rm He} \to ^4{\rm He}) = 297 \pm 19 {\rm mb} \;,\]
\[ \sigma^{(I)}(^8{\rm He}) - \sigma^{(I)}(^4{\rm He}) = 314 \pm 8 {\rm mb} \;.\]
\end{center}

\begin{table}
\caption{\label{tab:table4}Production cross sections of the projectile fragments $\sigma(AB \to F+X)$ on $^{12}$C target at 790 MeV/nucleon and differences of beam and fragment reaction cross sections also on $^{12}$C target.}
{\begin{tabular}{@{}cccc@{}} \toprule
Beam & Fragment & $\sigma(AB \to F+X)$ & $\sigma^{(r)}_{AB} - \sigma^{(r)}_{FB}$ \\ 
 & & (mb) & (mb) \\ \colrule
$^{14}$Be & $^{12}$Be  & $210 \pm 10$ \footnotemark[1] & $182 \pm 71$ \footnotemark[1]\\  
$^{11}$Be & $^{10}$Be  & $169 \pm 4$ \footnotemark[1]  & $129 \pm 13$ \footnotemark[1]\\   
$^{11}$Li & $^9$Li     & $213 \pm 21$ \footnotemark[2]  & $260 \pm 20$ \footnotemark[1]\\
$^8$He    & $^6$He     & $202 \pm 17$ \footnotemark[2]  &  $95 \pm 9$ \footnotemark[1]\\
$^8$He    & $^4$He     & $95  \pm 9$ \footnotemark[2]  & $314 \pm 8$ \footnotemark[1]\\
$^6$He    & $^4$He     & $189 \pm 14$ \footnotemark[2] & $219 \pm 8$ \footnotemark[1]\\   \botrule
\end{tabular}}
\footnotetext[1]{Data were taken from \cite{Tani8}.}
\footnotetext[2]{Data were taken from \cite{Koba}.}
\end{table}

\setcounter{equation}{0}
\section{\label{sec:section5}Matter and charge rms radii in unstable nuclei}

\subsection{\label{sec:section5.1}Data at low energies}
There is a large volume of the experimental data on nucleus-nucleus interactions obtained at relatively low energies smaller than 300-400 MeV per nucleon. These data can be useful for measurements of the mass of nuclide, energy separation of a valence nucleon, etc.

The data shows the qualitative difference in differential cross sections of elastic $p\,^9$Li and  $p\,^{11}$Li scattering \cite{Moon}. That provides additional arguments for an existence of neutron halo in $^{11}$Li nucleus. However, independently on the quality of available data, the quantitative interpretation of the data and the extraction of parameters of the nuclear density distribution (or R$_{\rm m}$) for unstable nuclides are problematic.

As it was mentioned in Section~\ref{sec:section2.1}, the corrections to the interaction and reaction cross sections calculated in the framework of  Glauber Theory become significant in the low-energy range.  For example, it was shown in \cite{Take} that taking into account a Fermi-motion of nucleons in projectile and target nuclei changes the effective energy of nucleon-nucleon interaction ($NN$). Since the cross section of $NN$ interaction has a strong dependence on the effective energy, when at energies below 100 MeV per nucleon, the correction to the reaction cross section $\sigma^{(r)}$ reaches 10\% of its value.

At smaller energies the modification of the fast proton trajectories by the Coulomb field leads to even more significant corrections. It was shown in \cite{Shuk} that at energy 30 MeV per nucleon this correction can change value of $\sigma^{(r)}$ for $^{12}$C - $^{40}$Ca collision by approximately 10\%.  Moreover, the mentioned above corrections are significant even  for calculations of low-energy proton-nucleus inelastic cross sections, \cite{DDP}. 

In order to extract parameters of nuclear density distribution the following semiemprirical parameterization of reaction (or interaction) cross sections was suggested in \cite{Kox}:
\begin{equation}
\label{5.1}
\sigma^{(I)}  = \pi R^2_0 \left[A^{1/3}_p +  A^{1/3}_t + a \frac{A^{1/3}_p  A^{1/3}_t} {A^{1/3}_p +  A^{1/3}_t} -C(E) \right]^2 \left[1-\frac{B_c}{E_{c.m.}}\right] \;.
\end{equation}
Here $R_0$ is radius of nucleon-nucleon interaction, $A_p$ and $A_t$ are the atomic weights of projectile and target nuclei, $a$ is an asymmetry parameter, $C(E)$ is the energy-dependent transparency parameter which is constant at $E  > 100$ MeV, and 
\begin{equation}
\label{5.2}
B_c = \frac{Z_pZ_te^2}{1.3(A^{1/3}_p +  A^{1/3}_t)}
\end{equation}
is the Coulomb barrier.

It was shown in \cite{Kox} that expression for the reaction (interaction) cross sections Eq.~(\ref{5.1}) is in good agreement with the experimental data obtained with stable nuclei. However, contrary to the general belief, this expression cannot be applied for the  quantitative analyses of experimental data with nuclei near the drip lines, where nuclei with the halo or skin can exist, \cite{Tani8}.

First of all, the parameter $C(E)$ which accounts for the diffuseness of nuclear surface (similarly to the parameter $c$ in Eq.~(\ref{1.5})) obtained from the analysis of interaction between stable nuclei cannot be used to describe data obtained in experiments with unstable nuclei. Clearly, the value for the diffuseness parameter for unstable nuclei with halo should be significantly higher compare to its value for stable nuclei. Secondly, semiempirical expression Eq.~(\ref{5.1}) suggests that the the value of $R_0$ is the same for projectile and target nuclei, which is inconsistent with the experimental data obtained at low energy \cite{Mitt}.  For example, the based on the Eq.~(\ref{5.1}) analysis would yield different values for the radius of the target nucleus $^{12}$C when applied for the data obtained in $^{12}$C-$^{12}$C and $^{11}$Li-$^{12}$C collisions. As a result, anomalously large value for the radius of $^{11}$Li nuclide becomes "distributed" between the radii of  $^{11}$Li and $^{12}$C.  

For these reasons we believe that expression Eq.~(\ref{5.1}) cannot be used to analyze experimental data with radioactive nuclei, and, hence, we will not present the values of $\rm R_m$ obtained with the help of this expression in Table~\ref{tab:table5} in next section.

Of course, all these criticism cannot be applied to the case of qualitative comparison of two projectile nuclides interacting with the same target. 

\subsection{\label{sec:section5.2}Summary of results for nuclear radii extracted from interaction cross section data at high energies}

In this section we present the values for root-mean-square radii, R$_{\rm m}$, of nuclear matter distributions extracted from the experimental data on  interaction cross sections obtained at somewhat high energies (higher than 600 MeV per nucleon). All data on rms radii presented in this section were obtained in optical approximation of the Glauber theory. Comparison with the data for nuclear radii presented in Table~1 shows that the difference between values for nuclear radius obtained using exact formulas of the Glauber Theory and values obtained in optical approximation is approximately the same for all nuclei. We do not present the values of interaction radii obtained by using Eq.~(\ref{1.4}) since these radii were obtained without taking into account the fact that the diffuseness region of the nuclear density distribution (parameter $c$ in Eq.~(\ref{1.5})) varies significantly when a halo present.

The data\footnote{When using these data, it is necessary to keep in mind the criticism presented in \cite{ElTa}.} presented in \cite{Tani1} were obtained using a shell-model harmonic oscillator function \cite{Elt} as a nuclear matter density distributions. The results obtained after an improvement of the data presented in \cite{Tani1} together with new results for Be and B isotopes are presented in \cite{Tani11}. In the later reference free values of $NN$ cross sections were used and the R$_{\rm m}$ values for point nucleon distributions are presented. Due to these reasons the R$_{\rm m}$ values presented in \cite{Tani11} are significantly smaller in comparison with the results in \cite{Tani1}. 

In the case of $^8$B nucleus there are some inconsistencies between somewhat small R$_{\rm m}$ value obtained using the harmonic oscillator density  distribution in \cite{Obu} and very small separation energy of one proton \cite{Mina}, $S_{1p}$ = 0.14 MeV, as well as with results of calculations,  see in \cite{Obu}. In \cite{Khal} $^8$B nucleus was considered as a possible candidate for having the proton halo, see Fig.~7. It is necessary to note that the  R$_{\rm m}$ value obtained in \cite{Obu} for $^8$Li nuclide was also smaller than the value presented in \cite{Tani1}, see Table~\ref{tab:table5}.

The values of R$_{\rm m}$ for carbon isotopes \cite{Oza4} as well as for $^{17}$N, $^{17}$F, and $^{17}$Ne were obtained for point-like nucleons using harmonic oscillator nuclear matter density distribution \cite{Oza3}. 

The nuclear matter radii of A=20 isobars were analyzed in \cite{Chul} using Woods-Saxon nuclear density distributions. Results of this analysis are also presented in Table~\ref{tab:table5}. 

\begin{table}[pt]
\caption{\label{tab:table5}R$_{\rm m}$ values of nuclear matter distributions extracted from the data on high energy ($E_{lab} >$ 600 MeV) nucleus-nucleus collisions.}
{\begin{tabular}{ccc||ccc} \toprule
Nucleus &  R$_{rm m}$ (fm) & Ref. & Nucleus &  R$_{rm m}$ (fm) & Ref. \\ \colrule
$^{4}$He   & $1.72 \pm 0.06$ & \cite{Tani1}  & $^{14}$Be  & $3.16 \pm 0.38$ & \cite{Tani11} \\    
$^{4}$He   & $1.57 \pm 0.04$ & \cite{Tani11}  & $^8$B     & $2.43 \pm 0.03$ & \cite{Obu}    \\    
$^{6}$He   & $2.87 \pm 0.04$ & \cite{Tani1} & $^8$B     & $2.38 \pm 0.04$ & \cite{Tani11}  \\    
$^{6}$He   & $2.48 \pm 0.03$ & \cite{Tani11} & $^{12}$B  & $2.39 \pm 0.02$  & \cite{Tani11}  \\   
$^{8}$He   & $2.81 \pm 0.03$ & \cite{Tani1} & $^{13}$B  & $2.46 \pm 0.12$  & \cite{Tani11}  \\   
$^{8}$He   & $2.52 \pm 0.03$ & \cite{Tani11} & $^{14}$B   & $2.44 \pm 0.06$ & \cite{Tani11} \\   
$^6$Li     & $2.54 \pm 0.03$ & \cite{Tani1} &  $^{15}$B   & $2.47 \pm 0.27$ & \cite{Tani11} \\   
$^6$Li     & $2.32 \pm 0.03$ & \cite{Tani11} & $^{12}$C   & $2.43 \pm 0.02$ & \cite{Tani1}  \\    
$^7$Li     & $2.54 \pm 0.03$ & \cite{Tani1} & $^{12}$C   & $2.31 \pm 0.02$ & \cite{Oza4} \\   
$^7$Li     & $2.33 \pm 0.02$ & \cite{Tani11} & $^{13}$C   & $2.28 \pm 0.04$ & \cite{Oza4}  \\   
$^8$Li     & $2.57 \pm 0.03$ & \cite{Tani1} & $^{14}$C   & $2.30 \pm 0.07$ & \cite{Oza4}  \\    
$^8$Li     & $2.37 \pm 0.02$ & \cite{Tani11} & $^{16}$C   & $2.70 \pm 0.03$ & \cite{Oza4}  \\   
$^8$Li     & $2.37 \pm 0.02$ & \cite{Obu} & $^{17}$C   & $2.72 \pm 0.03$ & \cite{Oza4}   \\   
$^9$Li     & $2.50 \pm 0.02$ & \cite{Tani1} & $^{18}$C   & $2.82 \pm 0.04$ & \cite{Oza4} \\    
$^9$Li     & $2.32 \pm 0.02$ & \cite{Tani11} & $^{19}$C   & $3.13 \pm 0.07$ & \cite{Oza4}  \\    
$^{11}$Li  & $3.36 \pm 0.24$ & \cite{Tani1}  & $^{20}$C   & $2.98 \pm 0.05$ & \cite{Oza4}  \\    
$^{11}$Li  & $3.12 \pm 0.16$ & \cite{Tani11} & $^{17}$N   & $2.48 \pm 0.05$ & \cite{Oza3} \\  
$^7$Be     & $2.41 \pm 0.03$ & \cite{Tani1} & $^{20}$N   & $2.77 \pm 0.04$ & \cite{Chul}  \\    
$^7$Be     & $2.31 \pm 0.02$ & \cite{Tani11} & $^{20}$O   & $2.64 \pm 0.03$ & \cite{Chul}   \\  
$^9$Be     & $2.53 \pm 0.01$ & \cite{Tani1} & $^{17}$F  & $2.54 \pm 0.08$ & \cite{Oza3}   \\   
$^9$Be     & $2.38 \pm 0.01$ & \cite{Tani11} & $^{20}$F   & $2.75 \pm 0.03$ & \cite{Chul}  \\   
$^{10}$Be  & $2.43 \pm 0.02$ & \cite{Tani1} & $^{17}$Ne  & $2.75 \pm 0.07$ & \cite{Oza3}  \\   
$^{10}$Be  & $2.30 \pm 0.02$ & \cite{Tani11} & $^{20}$Ne  & $2.84 \pm 0.03$ & \cite{Chul}  \\   
$^{11}$Be  & $2.73 \pm 0.05$ & \cite{Tani11} & $^{20}$Na  & $2.69 \pm 0.03$ & \cite{Chul}  \\   
$^{12}$Be  & $2.59 \pm 0.06$ & \cite{Tani11} & $^{20}$Mg  & $2.86 \pm 0.03$ & \cite{Chul}  \\   \botrule
\end{tabular}}
\end{table}

\subsection{\label{sec:section5.3}Charge radii of light nuclei}

The classical way to measure the charge (or proton) radii of the nuclei is to analyze experimental data obtained in the eA elastic scattering experiments \cite{Elt1}. However, in the case of unstable nuclei this way is currently still under discussion \cite{Bert1,Wan1}. 

The best information about charge radii of unstable nuclei comes from the laser spectroscopy experiments. The energy levels (or the isotope shifts) of atomic electrons can be measured with very high accuracy, $\sim 10^{-6}$ or better, \cite{Dra,PuPa} in relative units. Experimental results are in very good agreement with theoretical calculations.

The isotope shifts between the atomic transitions in basic stable nuclei, $^{4}$He, $^7$Li and $^9$Be and their unstable isotopes are analyzed. Some contributions to the calculated values of the shifts are proportional to the ratio of the finite charge radii of the stable nucleus and its unstable isotope \cite{Dra,PuPa}. That allows to extract the difference in the values of $\rm R_{ch}$ in the basic stable nucleus and an isotope nuclide by using the results of theoretical calculations based on QED. The obtained results are presented in Table~\ref{tab:table6}. 
\begin{table}[pt]
\caption{\label{tab:table6}$\rm R_{ch}$ values of nuclear charge distributions obtained from laser spectroscopy data.}
{\begin{tabular}{cccc} \toprule
Nucleus   &  R$_{ch}$ (fm)     & Ref. & Mean life     \\  \colrule
$^{4}$He  & $1.673 \pm 0.001$  &  \cite{BRi}  & Stable    \\    
$^{6}$He  & $2.054 \pm 0.014$  & \cite{Wan} & 807 ms   \\  
$^{6}$He  & $2.068 \pm 0.011$  & \cite{Mue} &  -       \\  
$^{8}$He  & $1.93 \pm 0.03$    & \cite{Mue} & 119 ms   \\  
$^6$Li    & $2.51 \pm 0.06$    & \cite{Ewa} &  Stable   \\    
$^6$Li    & $2.540 \pm 0.028$  & \cite{PuPa} &  Stable   \\    
$^6$Li    & $2.49 \pm 0.04$    & \cite{Busha} &  -    \\    
$^6$Li    & $2.55 \pm 0.04$    &  \cite{DeJa}   &  -   \\    
$^7$Li    & $2.39 \pm 0.03$    &  \cite{DeJa}   & Stable   \\   
$^8$Li    & $2.29 \pm 0.08$    &  \cite{Ewa}   & 840 ms  \\    
$^8$Li    & $2.281 \pm 0.032$  &  \cite{PuPa}   & -  \\    
$^9$Li    & $2.22 \pm 0.09$    &  \cite{Ewa}   & 178 ms  \\    
$^9$Li    & $2.185 \pm 0.033$  &  \cite{PuPa}   & -  \\    
$^9$Li    & $2.217 \pm 0.035$  &  \cite{FCN,San}  & -  \\   
$^{11}$Li & $2.426 \pm 0.034$  &  \cite{PuPa}   & 8.6 ms  \\    
$^{11}$Li & $2.467 \pm 0.037$  &  \cite{San}   & -  \\    
$^7$Be     & $2.645 \pm 0.014$ &  \cite{PuPa}   &  53.2 d \\    
$^7$Be     & $2.647 \pm 0.017$ &  \cite{Nor}  & -   \\    
$^9$Be     & $2.519 \pm 0.012$ &  \cite{Nor}  & Stable   \\   
$^{10}$Be  & $2.357 \pm 0.018$ &  \cite{Nor}  & $1.5\times 10^6$ y  \\  
$^{11}$Be  & $2.463 \pm 0.016$ &  \cite{Nor}  & 13.8 s          \\  \botrule
\end{tabular}}
\end{table}

The radius of the point-like proton distribution, $\rm R_p$ was estimated in \cite{Wan} to be
\begin{equation}
\label{5.3}
\rm R_p(^6{\rm He}) = 1.912 \pm 0.018 \; \rm fm \;. 
\end{equation} 

One can see noticeable decrease of $\rm R_{ch}$ in the intervals from $^6$Li to $^9$Li and from $^7$Be to $^{11}$Be and $^{10}$Be and significant increase of  charge radii for $^{11}$Li and $^{11}$Be. That confirms the unusual structure of the last nuclides. The $^6$He, $^8$He, $^{11}$Li and  $^{11}$Be nuclei are supposed to consist of the $^4$He, $^4$He, $^9$Li and $^{10}$Be cores and halos of 2 neutrons, 4 neutrons, 2 neutrons and  1 neutron, respectively. However, the charge radii of $^6$He, $^8$He, $^{11}$Li and $^{11}$Be are larger than the corresponding radii of $^4$He, $^4$He, $^9$Li and $^{10}$Be (see data provided in Table~\ref{tab:table6}). The larger charge radii of  $^6$He, $^8$He, $^{11}$Li and $^{11}$Be can be explained by the core motion  around the center-of-mass of these nuclei and partially by the possible nuclear core polarization \cite{EHMS,Shul,Shimo,Alk1,Bertul,Ikeda}. The contribution to the effective core size due to its motion around the nuclear center-of-mass can be determined from the data on the Coulomb nuclear dissociation \cite{Suzu,Aumann,Alk2,Naka1,Naka2,Hagi} 

Some results on nuclear charged radii were extracted from the data on only interaction cross section. Since some additional assumptions on the nuclear structure are required to analyze experimental data, we will not present them in the article.

\setcounter{equation}{0}
\section{\label{sec:section6}Nuclear matter radii from proton-nucleus elastic scattering at intermediate energy}

Proton-nucleus elastic scattering at intermediate energy is an efficient means for studying nuclear matter distributions \cite{ABV}. There is an obvious advantage of proton scattering experiments at intermediate energy as compared to similar experiments at low energy. As we discussed before, at intermediate energy the mechanism of proton-nucleus scattering is somewhat simple and can be described in the framework of various modifications of multiple scattering theories, in particular the Glauber theory \cite{Gl2}. That allows one to connect rather accurately the measured differential scattering cross sections with the nuclear matter distributions under study. A number of experiments on proton-nucleus scattering were performed previously to study the matter distributions in stable nuclei. Such experiments can be also carried out to study matter distributions in short-lived unstable nuclei. In this case the experiments should be performed in inverse kinematics. 

The scattering of protons from the nuclear halo is confined to small scattering angles. Therefore, in order to study the spatial structure of halo nuclei, it is important to measure the differential cross sections for proton scattering at small momentum transfers. The analysis of the differential cross sections for proton scattering at small momentum transfers provided in \cite{Alk-Lob1,Alk-Lob2,Alk-Lob3} show that it is possible to determine both sizes of the nuclear core and halo. For the first time, the relevant experiments \cite{Al1,Neum,Dobr} were performed at GSI Darmstadt with the help of the ionization chamber IKAR \cite{Vorob} developed at PNPI Gatchina. 

\subsection{\label{sec:section6.1}Experiment}

Differential cross sections d$\sigma$/d$t$ were measured in inverse kinematics at GSI Darmstadt for proton scattering on nuclei of $^{4,6,8}$He and $^{6,8,9,11}$Li isotopes \cite{Neum,Dobr}. The measurements were performed at the equivalent proton energy $E_{\rm p} \approx$ 0.7 GeV, the range of the momentum transfer squared $t$ being $0.002 \leq |t| \leq 0.05$ (GeV/$c)^2$. A primary $^{18}$O beam from the heavy-ion synchrotron SIS at GSI Darmstadt was focused on an 8 g/cm$^2$ beryllium production target at the entrance of the fragment separator FRS. The helium and lithium ions, produced by fragmentation of $^{18}$O nuclei, were separated by the FRS according to their magnetic rigidity. The intensity of the secondary He and Li beams was about 10$^3$ s$^{-1}$ with a duty cycle in the range of 25 - 50 \%.

\begin{figure}[th]
\includegraphics[width=0.8\textwidth]{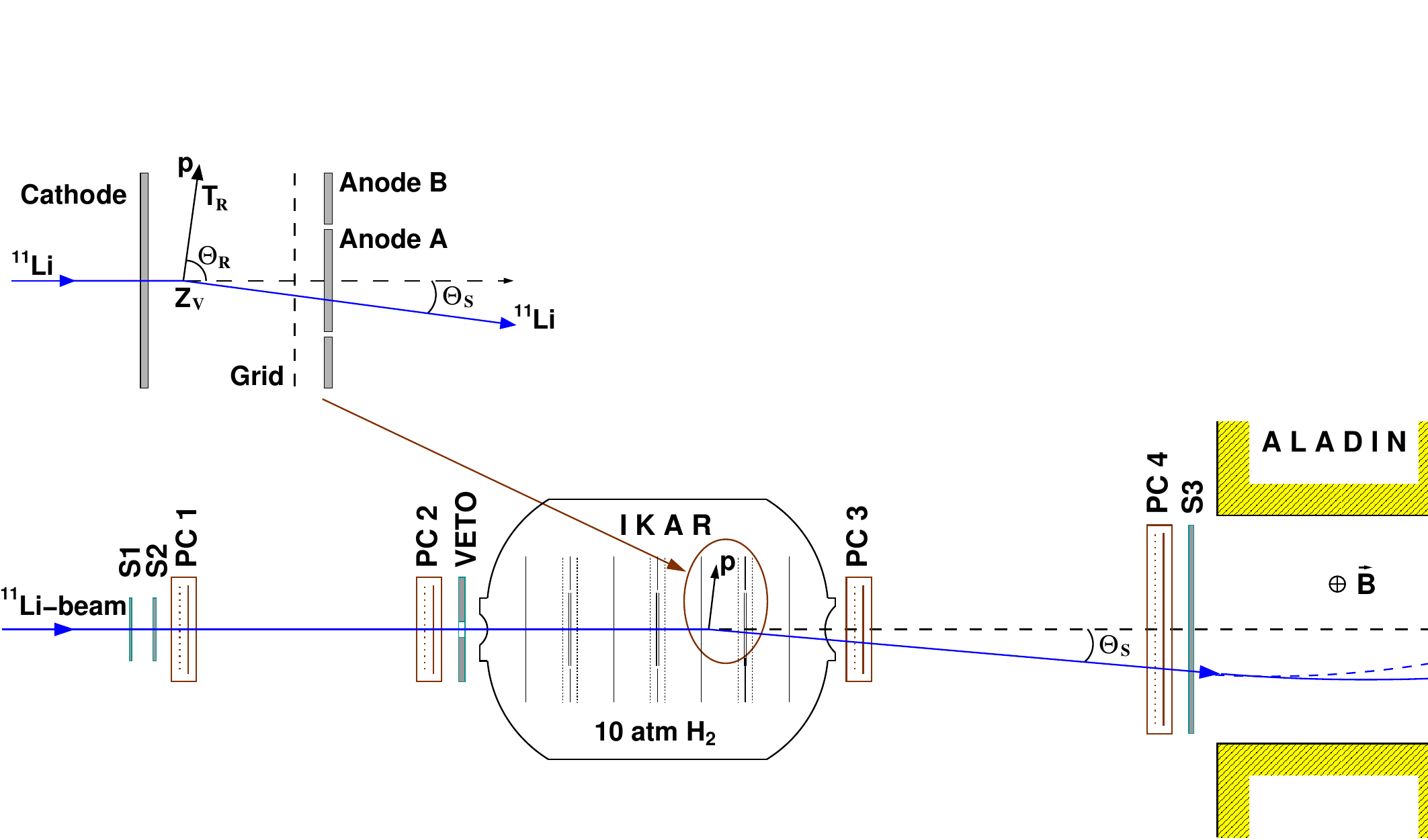}
\caption{Layout of the experimental setup for small-angle proton elastic scattering on exotic nuclei in inverse kinematics. IKAR -- hydrogen-filled ionization chamber  which serves as gas target and detector of recoil protons. PC1-PC4 -- multi-wire proportional chambers, which measure the projectile scattering angle. S1-S3 and VETO -- scintillation counters for triggering and for beam particles identification. The ALADIN magnet with a position scintillator wall behind allows to identify the scattered beam particle and to discriminate the breakup channels.}
\end{figure}

A scheme of the layout of the experiments is shown in Fig. 9. The main component of the setup is the hydrogen-filled ionization chamber IKAR, which served as a gas target and simultaneously as a recoil proton detector. IKAR was operated at 10 bar pressure, which insured the effective H$_2$ thickness of about $3 \times 10^{22}$ protons/cm$^2$. IKAR consists of six independent identical modules. Each module is an axial ionization time-projection chamber, which contains anode plates, a cathode plate, and a grid (see insert in Fig. 9), all electrodes being arranged perpendicular to the beam direction. The signals from the electrodes provided the energy of the recoil proton (or its energy loss in case it leaves the active volume), the scattering angle of the recoil proton, and the coordinate along the chamber axis of the interaction point in the grid-cathode space. 

The recoil protons in IKAR were registered in coincidence with the scattered He or Li particles. The momentum transfer could be determined either from the measured energy of the recoil proton or from the value of the projectile scattering angle, which was measured by a tracking detector consisting of 2 pairs of two-dimensional multi-wire proportional chambers arranged upstream and downstream with respect to IKAR (see Fig. 9). A set of scintillation counters was used for triggering and identification of the beam particles via time-of-flight and d$E$/d$x$ measurements, while a circular-aperture scintillator VETO selected the projectiles which entered IKAR within an area of 2 cm in diameter around the central axis. 

In the case of the experiment with Li isotopes \cite{Dobr} a magnetic-rigidity analysis of the scattered particles was also performed with the help of a large-gap magnet (ALADIN) and a scintillator wall behind it, which allowed one to exclude a contribution from the break-up channels. The systematic uncertainty in the normalization of the measured cross sections was estimated to be about 3\%, while the uncertainty in the $t$-scale calibration was about 1.5 \%. 

\begin{figure}[th]
\includegraphics[width=0.7\textwidth]{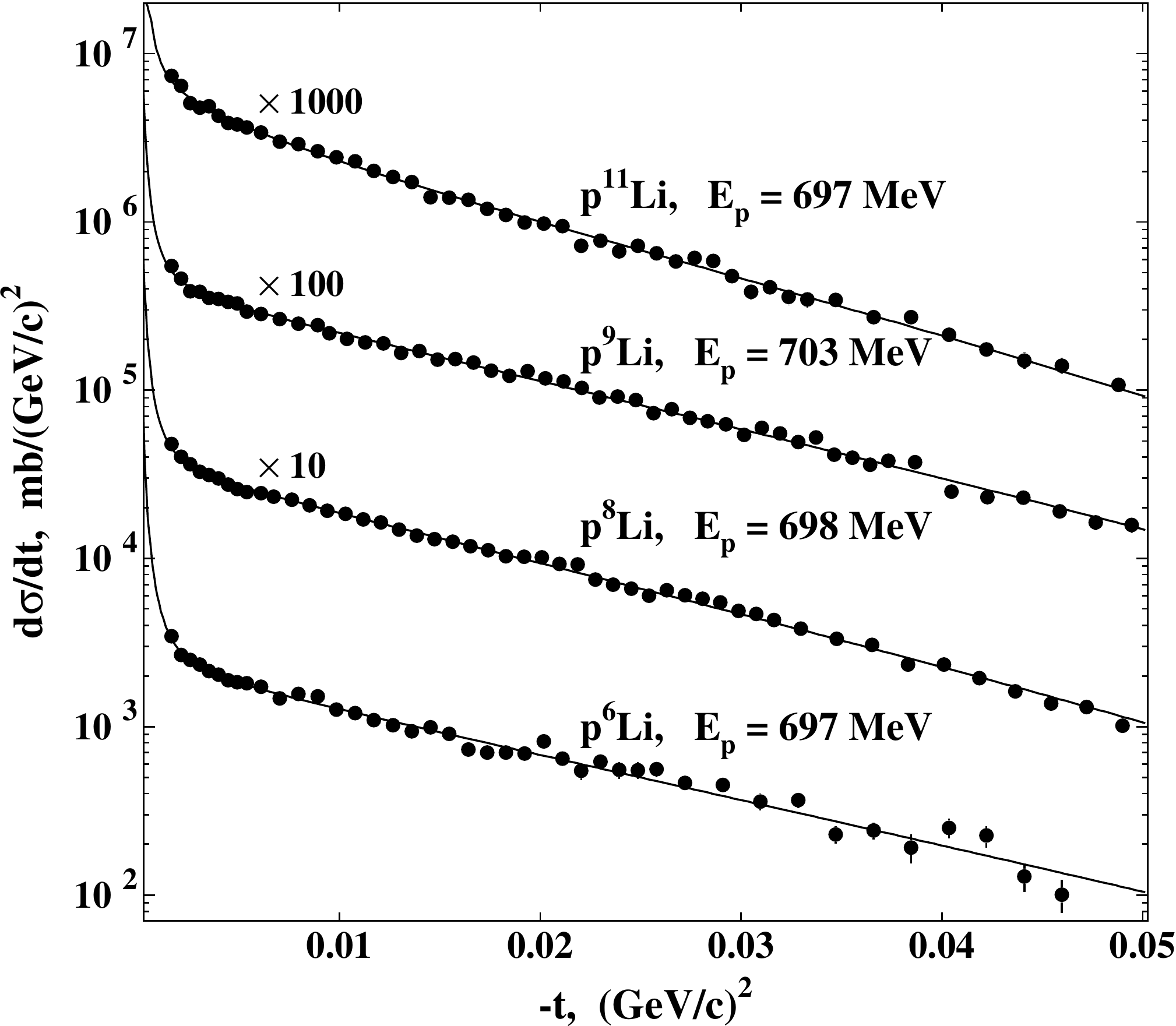}
\caption{Measured differential cross sections for $p^{6,8,9,11}$Li elastic scattering (dots) versus the four-momentum transfer squared. Solid lines are the cross sections calculated within the Glauber theory using the GO parameterization for the matter density distributions with the fitted parameters.}
\end{figure}

The differential cross sections d$\sigma$/d$t$ measured for the case of Li isotopes is shown in Fig. 10. At first glance all cross sections have similar behavior. A steep rise of the cross sections with $|t|$ decreasing at $|t| \leq 0.004$ (GeV/$c)^2$ is due to the Coulomb scattering. At $|t| > 0.005$ (GeV/$c)^2$, the cross sections decrease with $|t|$ approximately as exponents. However, if one divides the cross sections by exponents (see in Fig. 11), then it is seen that the dependence of d$\sigma$/d$t$ on $|t|$ for $p^{6,8,9}$Li and for $p^{11}$Li scattering is different. The shape of d$\sigma$/d$t$ for the case of proton scattering on $^{6,8,9}$Li nuclei (at $0.005 \leq |t| \leq 0.05$ (GeV/$c)^2$) is indeed close to that of exponents, whereas the shape of d$\sigma$/d$t$ for $p^{11}$Li scattering deviates significantly from the exponential one. As will be discussed in Section~\ref{sec:section6.2}, such a shape of the cross section is an indication for the core $+$ halo structure of the investigated nucleus. 

\begin{figure}[th]
\includegraphics[width=0.7\textwidth]{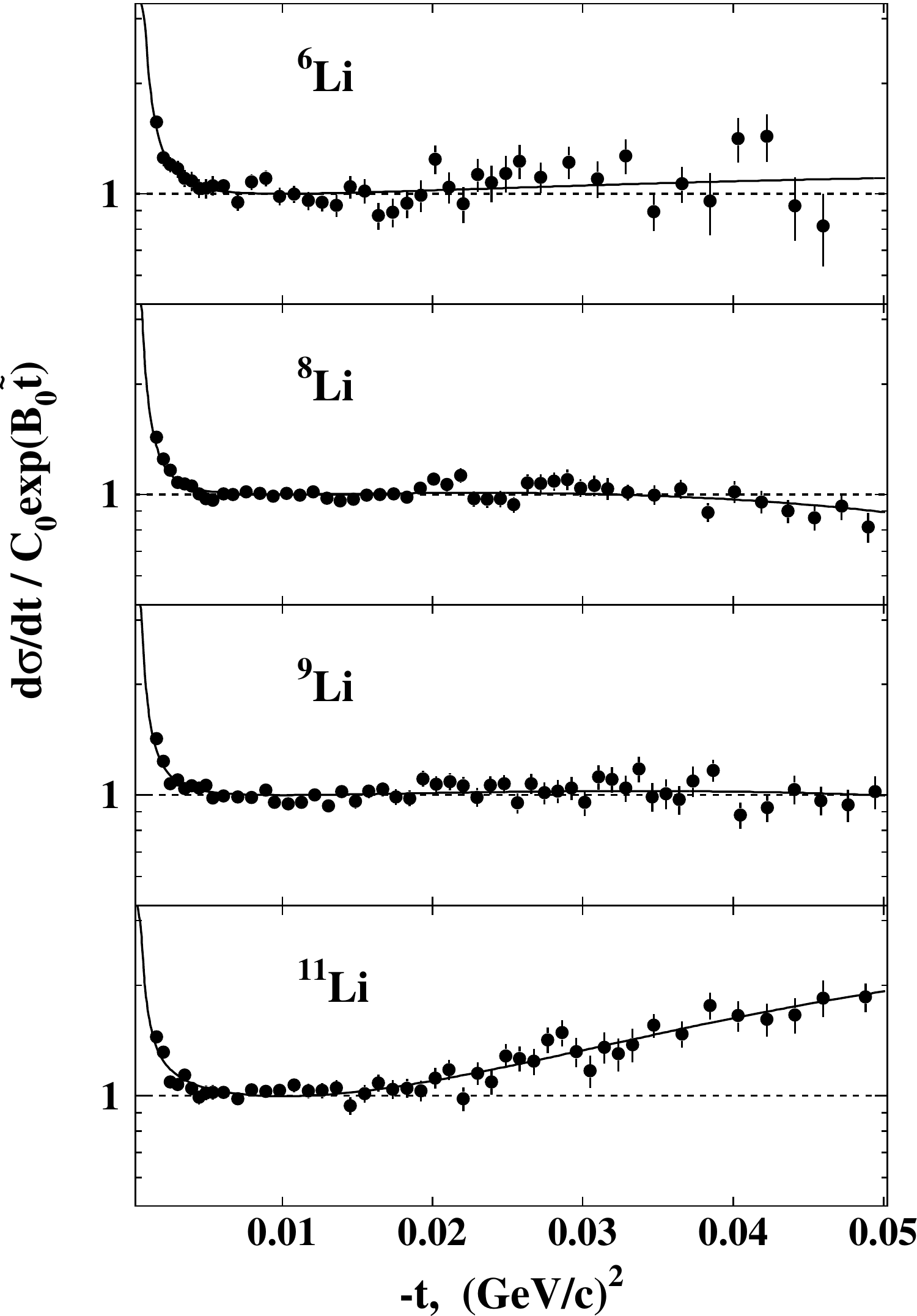}
\caption{Differential cross sections, the same as in Fig. 10, divided by exponents as explained in the text.}
\end{figure}

\subsection{\label{sec:section6.2}Analysis of the data}

The differential cross sections for proton elastic scattering on the studied nuclei were calculated by the Glauber formula (Eq. 2.8) using phenomenological density distributions, each having two free parameters, which were determined by fitting the calculated cross sections to the experimental data. As we discussed in Section~\ref{sec:section2}, the $pN$ scattering amplitude can be described by the standard high-energy parameterization
\begin{equation}
\label{6.1}
f_{\rm pN}({\bf q}) = ({\rm i}k/4\pi)\sigma_{\rm pN}(1 - {\rm i}\epsilon_{\rm pN}){\rm exp}( - \beta_{\rm pN}{\bf q}^2 /2) \; ,
\end{equation}
where $\sigma_{\rm pp}$ and $\sigma_{\rm pn}$ are the total cross sections of $pp$ and $pn$ interaction, $\epsilon_{\rm pp}$ and $\epsilon_{\rm pn}$ are the ratios of the real to imaginary parts, and $\beta_{\rm pp}$ and $\beta_{\rm pn}$ are the slope parameters. The slope parameters $\beta_{\rm pp}$ and $\beta_{\rm pn}$ were evaluated from the experimental data and partial wave analyses for free $pp$ and $pn$ scattering. 

Four parameterizations for phenomenological nuclear density distributions were applied, labeled as SF (symmetrized Fermi), GH (Gaussian-halo), GG (Gaussian-Gaussian) and GO (Gaussian-oscillator). In the SF parameterization \cite{Eldysh}, the free parameters are the "half density radius" $R_{\rm 0}$ and the diffuseness parameter $a$. The GH parameterization \cite{Al1} is defined by the formfactor 
$$
S(t) = (1 + \alpha z^2) {\rm exp}(z) \; ,
$$
where $R_{\rm m}$ is the root-mean-square radius of the matter distribution and 0 $\leq \alpha \leq$ 0.4 and $z = t R^2_{\rm m}/6$. The GH distribution becomes a Gaussian one when $\alpha = 0$, whereas for $\alpha$ close to 0.4 this distribution has a pronounced halo component. 

While the SF and GH parameterizations do not set apart the neutron and proton distributions, the GG and GO parameterizations assume that the nuclei consist of core nucleons and valence nucleons with different spatial distributions. The core nucleon distribution is assumed to be a Gaussian one in both the GG and GO parameterizations. The valence nucleon density is described by a Gaussian or a 1$p$ shell harmonic oscillator type distribution within the GG or GO parameterizations, respectively. 

The free parameters in the GG and GO parameterizations are the rms radii $R_{\rm c}$ and $R_{\rm v}$ of the core and valence nucleon distributions. The explicit expressions for the SF, GH, GG, and GO parameterizations are given in \cite{Al2}. The studied nuclei were considered to have one, two, or four valence nucleons. The cores in $^6$He, $^8$He, $^6$Li, $^8$Li, $^9$Li, and $^{11}$Li were presumed to have the nucleon composition and the spatial structure similar to that of the $^4$He, $^4$He, $^4$He, $^7$Li, $^7$Li, and $^9$Li nuclei, respectively. 

\begin{figure}[th]
\includegraphics[width=0.7\textwidth]{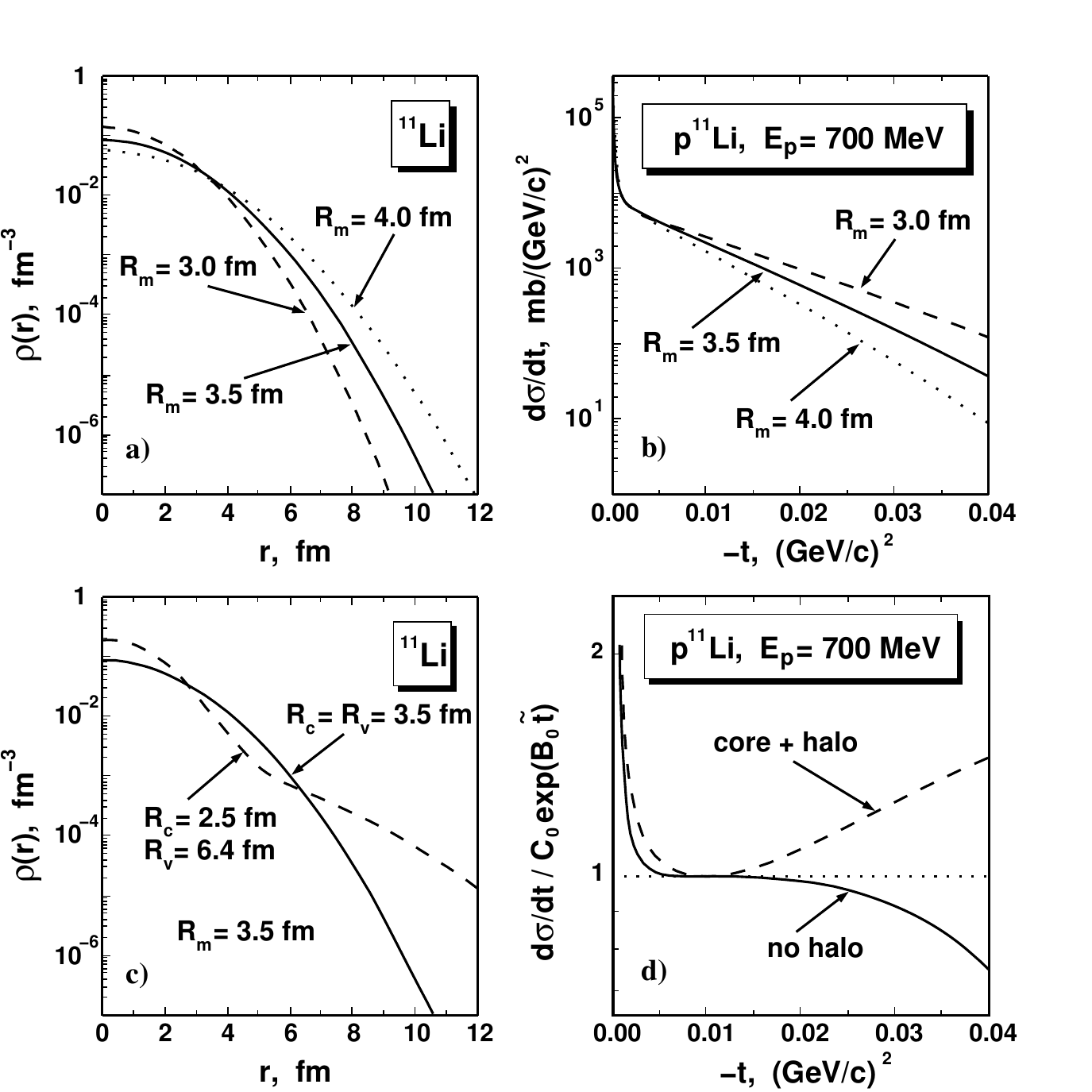}
\caption{The sensitivity of the differential cross sections for small-angle proton elastic scattering to the nuclear size and the radial shape of the nuclear matter distribution.}
\end{figure}

The sensitivity of the different cross sections to the nuclear size and radial shape of the nuclear matter distribution is demonstrated in Fig. 12, where results of calculations for $p^{11}$Li scattering at 0.7 GeV are displayed. The differential cross sections d$\sigma$/d$t$ for small-angle elastic scattering is shown in Fig. 12 (b) as a function of $t$. They were calculated for a Gaussian matter distribution $\rho (r)$ with different matter radii $R_{\rm m}$ (see Fig. 12 (a)). A strong correlation between the slope of the cross section and the radius $R_{\rm m}$ is obvious. The lower part of Fig. 12 demonstrates the sensitivity of the calculated cross section to the radial shape of the nuclear matter distribution. Two different nuclear matter density distributions are assumed, one being a Gaussian with $R_{\rm m} = R_{\rm c} = R_{\rm v} = 3.5$ fm, the other being the sum of two Gaussians -- one for the core nucleons (with $R_{\rm c} = 2.5$ fm) and the other for the valence (halo) nucleons ($R_{\rm v}= 6.4$ fm). Both density distributions (shown in Fig. 12 (c)) have the same nuclear matter radius $R_{\rm m}$ but significantly different radial shapes. Fig. 12 (d) depicts in the logarithmic scale the calculated cross sections related to these densities. In order to see more clearly the sensitivity of the cross sections to the density shape, the calculated cross sections are normalized by the exponential functions $C_0 \times $exp($B_0\tilde{t}$). Here $\tilde{t} = t - t^*$, $t^* = - 0.01$ (GeV/$c$)$^2$, while the quantities $C_0$ and $B_0$ are the values of the differential cross sections and their slope parameters at $t = t^*$, corresponding to the two density distributions involved in the calculations. It is seen that the shapes of the cross sections is significantly different for the two cases considered, the shape calculated assuming the core+halo structure of $^{11}$Li being similar to that of the experimental cross section shown in Fig. 11.

The results of the data analysis using phenomenological density distributions are presented in Table~\ref{tab:table7}. In the cases of $^{4,6,8}$He and $^{6,8,9}$Li nuclei, good data fit was achieved for all four density parameterizations applied. However, in the case of $^{11}$Li, only GG and GO parameterizations, which allow different distributions of the core and valence (halo) nucleons, permitted good data description.

\begin{table}[pt]
\caption{\label{tab:table7}Root-mean-square radii $R_{\rm c}$, $R_{\rm v}$ and $R_{\rm m}$ correspondingly of the core nucleon, valence (halo) nucleon and total matter distributions, deduced from the data on small-angle proton elastic scattering.}
{\begin{tabular}{@{}cccc@{}} \toprule
 Nucleus   &  $R_{\rm c}$ (fm)   & $R_{\rm v}$ (fm) & $R_{\rm m}$ (fm) \\ \colrule
 $^4$He    &  ---               &  ---            & 1.49(3)         \\
 $^6$He    &  1.88(12)          & 2.97(26)        & 2.30(7)         \\
 $^8$He    &  1.55(15)          & 3.08(10)        & 2.45(7)         \\
 $^6$Li    &  2.10(15)          & 3.00(32)        & 2.44(7)         \\
 $^8$Li    &  2.48(7)           & 2.58(48)        & 2.50(6)         \\
 $^9$Li    &  2.20(6)           & 3.12(28)        & 2.44(6)         \\
 $^{11}$Li &  2.52(2)           & 5.98(32)        & 3.42(11)        \\ \botrule
\end{tabular}}
\end{table}

The obtained results can be interpreted as an indication on the core+halo (skin) structure of $^6$He, $^8$He, $^6$Li, $^8$Li, $^9$Li and $^{11}$Li. Within the quoted errors the matter radii of the $^6$Li, $^8$Li and $^9$Li nuclei are identical. This means that $^8$Li and, especially, $^9$Li are more dense nuclei than $^6$Li. The latter is being considered to have an $\alpha + d$ spatial structure. The determined core and halo radii represent clear evidence of a neutron halo in $^{11}$Li. Indeed, the deduced halo radius $R_{\rm h} \equiv R_{\rm v} = 5.98(32)$ fm is larger than the core radius $R_{\rm c}$ = 2.52 fm by a factor of more than 2. The total matter radius $R_{\rm m}$ of $^{11}$Li is significantly larger than those of the lighter Li isotopes. This result being in agreement with the data on nucleus-nucleus interaction cross sections discussed before. 

Among the studied nuclei of He and Li isotopes, $^8$He and $^{11}$Li have the most developed halo-like (skin-like) structure. In Fig. 13 (a) and (b), the core and total matter density distributions derived for $^8$He and $^{11}$Li within the GG and GO parameterizations are compared. In the case of $^8$He it is still unclear if  the valence neutron distribution is halo or skin like. The observed extended valence neutron distribution in $^{11}$Li at the nuclear periphery is decidedly an outstanding halo. 

\begin{figure}[th]
\includegraphics[width=0.8\textwidth]{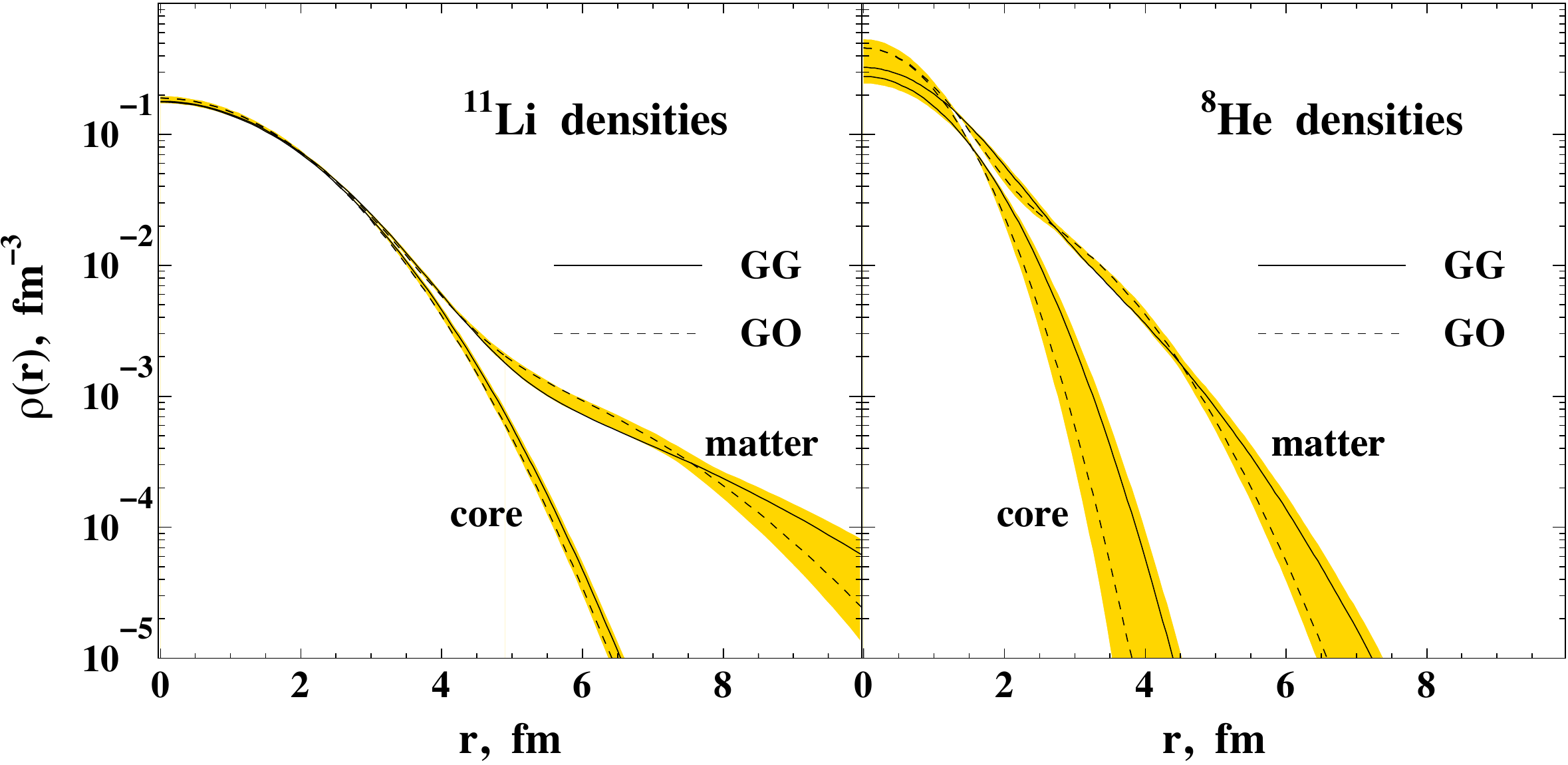}
\caption{Nuclear core and total matter distributions in $^8$He and $^{11}$Li, deduced from the cross sections for elastic $p^8$He and $p^{11}$Li scattering with the help of GG and GO density distribution parameterizations. The shaded areas represent the envelopes of the matter and core density variations within the model parameterizations used, superimposed by the statistical errors.}
\end{figure}

In the analysis of the cross sections with phenomenological densities discussed above all nucleon correlation (except the centre-of-mass correlations) in the nuclear many body density distributions were neglected. The deduced matter radii of $^6$He and $^8$He were $R_{\rm m}(^6$He) = 2.30 (7) fm and $R_{\rm m}(^8$He) = 2.45 (7) fm \cite{Al1}. However, the analysis of the same data carried out later by Al-Khalili and Tostevin in \cite{Al-Kha} using theoretical density distributions, the obtained radii of the same nuclei are larger by approximately 0.2 fm. Al-Khalili and Tostevin believed that larger values of $R_{\rm m}$ than those in \cite{Al1} were obtained because few-body correlations in $^6$He and $^8$He were treated properly  and used density distributions were with correct asymptotic. Authors stressed that the calculated cross sections and the deduced radii are somewhat sensitive to the few-body correlations in the many-body density and to the density asymptotic at large radii. This subject was considered in detail in \cite{Al2}. It was concluded in contrast to \cite{Al-Kha} that there is very weak sensitivity of the calculated cross sections (at small scattering angles) to the nuclear correlations and to the density asymptotic. On the other hand, the calculated matter radii $R_{\rm m}$ depend significantly on the density asymptotic. Theoretical density distributions \cite{Al-Kha} decrease with the radius increasing at large distances from the nuclear centre slower than the phenomenological density distributions in \cite{Al1}. We believe that provided reasons explain the larger values of $R_{\rm m}$ obtained in \cite{Al-Kha}. 

The nuclear density distributions in nuclei with low binding energies should have long density tails. Proton elastic scattering is sensitive to the nuclear spatial structure including the most part of the halo, however it is not sensitive to small density tails which contain only of the order of 1 percent (or even less) of the total matter. A contribution of such tails to the value of rms matter radii can be estimated theoretically. An analysis of the $p$He and $p$Li scattering cross sections with phenomenological density distributions including density tails taken from theoretical considerations was performed in \cite{Al2,Dobr}. The rms matter radii for $^6$He, $^8$He, and $^11$Li were found to be  $R_{\rm m}(^6$He) = 2.45 (10) fm, $R_{\rm m}(^8$He) = 2.53 (8) fm, and $R_{\rm m}(^{11}$Li) = 3.71 (20) fm. The relatively large error in the obtained value of $R_{\rm m}$ in the case of $^{11}$Li is mainly due to an uncertainty in the size and slope of the density distribution tail. A later analysis (not published) of the $p^{11}$Li scattering cross sections assuming a smaller contribution of the density tail yielded $R_{\rm m}(^{11}$Li) = 3.60 (20) fm.

Combining the matter radii $R_m$ obtained from the data on proton elastic scattering with the proton radii $R_{\rm p}$, determined from the nuclear charge radii measured in laser spectroscopy experiments (see Table~\ref{tab:table6}), the neutron radii $R_{\rm n}$ and the thickness of the neutron skin (halo) $\delta_{\rm np} = R_{\rm n} - R_{\rm p}$ can be determined. Values for $R_{\rm p}$, $R_{\rm n}$ and $\delta_{\rm np}$ are provided in Table~\ref{tab:table8}. 

\begin{table}[pt]
\caption{\label{tab:table8}Total matter radii $R_{\rm m}$ including contributions from density tails, neutron radii $R_{\rm n}$ deduced from matter radii $R_{\rm m}$ and the charge radii $R_{\rm ch}$, and the neutron skin (halo) thickness $\delta_{\rm np}$ in the $^{6,8}$He and $^{8,9,11}$Li nuclei.}
{\begin{tabular}{@{}cccc@{}} \toprule
 Nucleus   &  $R_{\rm m}$, fm   & $R_{\rm n}$, fm & $\delta_{\rm np}$, fm \\ \colrule
 $^6$He    &  2.45(10)          & 2.68(14)        & 0.76(14)              \\
 $^8$He    &  2.53(8)           & 2.73(10)        & 0.92(11)              \\
 $^8$Li    &  2.50(6)           & 2.68(9)         & 0.52(10)              \\
 $^9$Li    &  2.44(6)           & 2.59(9)         & 0.48(9)               \\
 $^{11}$Li &  3.60(20)          & 3.96(25)        & 1.58(25)              \\ \botrule
\end{tabular}}
\end{table}
\vskip 5pt

First experiments on the $p^{6,8}$He and $p^{8,9,11}$Li scattering in inverse kinematics \cite{Al1,Neum,Dobr} have shown that the intermediate-energy small-angle proton scattering is a useful means of investigation of the matter density distribution in light exotic nuclei. Future measurements of the cross sections for proton elastic scattering at larger momentum transfers \cite{Kis} will provide more detail information on the internal spatial structure of the studied nuclei. 

\setcounter{equation}{0}
\section{\label{sec:section7}Conclusion and Outlook}

There is an evident interest to the properties of radioactive nucleus. In nature there are  283 stable or very long lived nucleus \cite{Bert2} and about $(6 \div 8) \times 10^3$ radioactive nuclides \cite{Bert2,Ryj}. Only about half of the existing nuclei have been studied so far. The nuclear size is one of the basic parameter of the nuclear density distribution studying which is important for understanding of the nuclear properties. 

A lot of information on the matter radii has been obtained from the nucleus-nucleus interaction cross sections, which can be measured for very low intensity beams of exotic nuclei. However, the deduced matter radii are somewhat model dependent and are subject to some uncertainties appearing due to approximations used in the calculations of the reaction cross sections. The optical-limit approximation significantly overestimates the calculated reaction cross sections, especially in the case of halo nuclei. The rigid-target approximation provides somewhat more accurate results. In principle, the reaction cross sections can be expressed using the exact Glauber theory formulas and then numerically calculated using Monte-Carlo technique 

Provided the intensity of the nuclear beams is sufficient, the matter density distributions in exotic nuclei can be studied in intermediate-energy proton elastic scattering experiments in inverse kinematics. The charge radii of exotic nuclei are measured very precisely with the laser-spectroscopy technique. 

New experimental facilities for studying the properties of nuclei far from stability are planned to be built in the near future in Europe, Japan and the USA \cite{GANIL1,GANIL2,NuSTAR1,NuSTAR2,NuSTAR3,Riken1,Riken2,FRIB1,FRIB2}. Studying nuclei at these facilities will significantly increase our understanding of the spatial structure of the nuclei far from stability.

Ambitious project  NuSTAR at FAIR (Darmstadt, Germany) \cite{NuSTAR1,NuSTAR2,NuSTAR3}  will provide new fascinating possibilities for studying radioactive nuclei. New facility will produce intensive intermediate-energy and low-energy beams of nuclei far from stability which will allow to carry out versatile investigations of the nuclear properties. In particular, matter radii will be determined for long isotopic chains of many elements from the measured nucleus-nucleus interaction cross sections. Experiments on proton elastic scattering will be used to obtain information on matter density distributions. Collaboration also plan to use laser-spectroscopy technique to measure nuclear charge radii and electron elastic scattering experiments to determine nuclear charge distributions.

We expect that the improvement of the existing experimental technique and new facility will advance our understanding of the unstable nuclei.

\section*{Acknowledgements}

This work was supported in part by grant RSGSS--3628.2008.2.


\newpage

\begin{thebibliography}{0}

\bibitem{Elt} L. R. B. Elton. Proc. Phys. Soc. (London) {\bf A63} (1950) 1115.
\bibitem{Hof1} Electron Scattering, Nuclear and Nucleon Structure. Ed. by R.~Hofst\"adter. W. A. Benjamin, Inc. New-York (1963).
\bibitem{Hof2} R. Hofst\"adter. Rev. Mod. Phys. {\bf 28} (1956) 214.
\bibitem{Elt1} L.R.B. Elton, Nuclear Sizes. Oxford Univ. Press, New York (1961).
\bibitem{BP} H.L. Bradt and L. Peters. Phys. Rev. {\bf 77} (1950) 54.
\bibitem{Abd} E.O. Abdrakhmanov et al., Z. Phys. {\bf C5} (1980) 1.
\bibitem{Fesh} H. Feshbach, Phys. Rev. {\bf 84}, 1206 (1951).
\bibitem{Bush} M.P. Bush et al., Phys. Rev. {\bf C53}, 3009 (1996).
\bibitem{CLS} A. Chaumeaux, V. Layly, and R. Schaeffer, Ann. Phys. {\bf 116}, 247 (1978).
\bibitem{ABV} G.D. Alkhazov, S.L. Belostotsky, and A.A. Vorobyov, Phys. Rep. {\bf 42}, 89 (1978).
\bibitem{Gl1} R.J. Glauber, Phys. Rev. {\bf 100}, 242 (1955).
\bibitem{Sit1} A.G. Sitenko, Ukr. Fiz. Journal {\bf 4}, 152 (1959).
\bibitem{Gl2} R.J. Glauber. In "Lectures in Theoretical Physics". Eds. W.E. Brittin et al., New York (1959), vol.1, p.315.
\bibitem{DeDe} H. De Vries, C.W. De Jager, and C. De Vries, Atom. Data Nucl. Data Tables {\bf 36}, 495 (1987).
\bibitem{Tani} I. Tanihata et al., Phys. Lett. {\bf B160}, 380 (1985).
\bibitem{Al1} G.D. Alkhazov et al., Phys. Rev Lett. {\bf 78}, 2313 (1997).
\bibitem{Oza1} A. Ozawa et al., Nucl. Phys. {\bf A709} (2002) 60; {\bf A727}, 465 (2003).
\bibitem{Oza2} A. Ozawa et al., Nucl. Phys. {\bf A673} (2000) 411.
\bibitem{Tani1} I. Tanihata et al., Phys. Rev. Lett. {\bf 55}, 2676 (1985).
\bibitem{Al2} G.D. Alkhazov et al., Nucl. Phys. {\bf A712}, 269 (2002).
\bibitem{Kob} T. Kobayashi et al., Phys. Lett. {\bf B232}, 51 (1989).
\bibitem{HL} H. Heckman and P.J. Lindstrom, Phys. Rev. Lett. {\bf 37}, 56 (1976).
\bibitem{KoKo} V.M. Kolybasov and L.A. Kondratyuk. Phys. Lett. {\bf B39} (1972) 439.
\bibitem{LaLi} L.D. Landau and E.M. Lifshitz. Quantum Mechanics (Non-relativistic theory). Pergamon Press (1965).
\bibitem{LISchiff} L. I. Schiff. Quantum Mechanics (International Pure and Applied Physics Series). McGraw-Hill Companies (1968).
\bibitem{PaRa} C. Pajares and A.V. Ramallo, Phys. Rev. {\bf C16}, 2800 (1985).
\bibitem{BrSh} V.M. Braun and Yu.M. Shabelski, Int. J. of Mod. Phys. {\bf A3}, 2417 (1988).
\bibitem{CzMa} W. Czyz and L.G. Maximon, Ann. Phys. {\bf 52}, 59 (1969).
\bibitem{Karol} P.J. Karol, Phys. Rev. {\bf C11}, 1203 (1975).
\bibitem{FV} V. Franco and G.K. Varma, Phys. Rev. {\bf C18}, 349 (1978).
\bibitem{Sh2} Yu.M. Shabelski, Yad. Fiz. {\bf 47}, 1612 (1988); Sov. J. Nucl. Phys. {\bf 47}, 1021 (1988).
\bibitem{Alk} G.D. Alkhazov et al., Nucl. Phys. {\bf A220}, 365 (1977).
\bibitem{VT} R.D. Viollier and E. Turtschi, Ann. Phys. {\bf 124}, 290 (1980).
\bibitem{Abu} B. Abu-Ibragim and Y. Suzuki, Phys. Rev. {\bf C61}, 051601(R)  (2000); .{\bf C62}, 034608 (2000); 
\bibitem{AlLo} G.D. Alkhazov and A.A. Lobodenko, Yad. Fiz. {\bf 70}, 98 (2007); Phys. Atom. Nucl. {\bf 70}, 93 (2007).
\bibitem{Andr1} I.V. Andreev and A.V. Chernov, Yad. Fiz. {\bf 28}, 477 (1978); Sov. J. Nucl. Phys. {\bf 28}, 243 (1978).
\bibitem{Andr2} I.V. Andreev and L.A. Khein, Yad. Fiz. {\bf 28}, 1499 (1978); Sov. J. Nucl. Phys. {\bf 28}, 770 (1978).
\bibitem{MBra} M.A. Braun, Yad. Fiz. {\bf 45}, 1625 (1987); Sov. J. Nucl. Phys. {\bf 45}, 1008 (1987).
\bibitem{BoKa} K.G. Boreskov and A.B. Kaidalov, Yad. Fiz. {\bf 48}, 575 (1988); Sov. J. Nucl. Phys. {\bf 48}, 367 (1988).
\bibitem{ZUS} A.M. Zadorozhnyj, V.V. Uzhinsky, and S.Yu. Shmakov, Yad. Fiz. {\bf 39}, 1165 (1984); Sov. J. Nucl. Phys. {\bf 39}, 729 (1984).
\bibitem{Shm} S.Yu. Shmakov et al., Comp. Phys. Commun {\bf 54}, 125 (1989).
\bibitem{Sh11} D. Krpic and Yu.M. Shabelski, Z. Phys. {\bf C48}, 483 (1990).
\bibitem{Gar} F.A. Gareev et al., Yad. Fiz. {\bf 58}, 620 (1995); Phys. Atom. Nucl. {\bf 58}, 564 (1995).
\bibitem{Var} K. Varga et al., Phys. Rev. {\bf C66}, 034611 (2002).
\bibitem{Metro} N. Metropolis et al., J. Chem. Phys. {\bf 21}, 1087 (1953)
\bibitem{Hast} W.K. Hastings, Biometrika, {\bf 57}, 97 (1970).
\bibitem{Oza} A. Ozawa, T. Suzuki, and I. Tanihata, Nucl. Phys. {\bf A693}, 32 (2001).
\bibitem{KSh} D. Krpic and Yu.M. Shabelski, Yad. Fiz. {\bf 52}, 766 (1990); Sov. J. Nucl. Phys. {\bf 52}, 490 (1990).
\bibitem{Oga} Y. Ogawa, K. Yabana, and Y. Suzuki, Nucl. Phys. {\bf A543}, 722 (1992).
\bibitem{MNS} C. Merino, I.S. Novikov and Yu.M. Shabelski, 	arXiv:0907.1697v1 [nucl-th].
\bibitem{Jar} J. Jaros et al., Phys. Rev. {\bf C18}, 2273 (1978).
\bibitem{WoSa} R. D. Woods and D. S. Saxon. Phys. Rev. {\bf 95} (1054) 577.
\bibitem{Tani2} I. Tanihata et al., Phys. Lett. {\bf B287}, 307 (1992).
\bibitem{FG} V. Franco and R. J. Glauber. Phys. Rev. {\bf 142} (1966) 1195.
\bibitem{KIO} A. Kohama, K. Iida, and K. Oyamatsu. Phys. Rev. {\bf C78}, 061601 (2008).
\bibitem{KIO1} A. Kohama, K. Iida, and K. Oyamatsu. Phys. Rev. {\bf C72}, 024602 (2005).
\bibitem{AGK} V.A. Abramovsky. V.N. Gribov, and O.V. Kancheli. Yad. Fiz {\bf 18}, 595 (1973); Sov. J. Nucl. Phys. {\bf 18}, 308 (1973).
\bibitem{Shab} Yu.M. Shabelski. Nucl. Phys. {\bf B132}, 491 (1978).
\bibitem{ShTr} Yu.M. Shabelski and D. Treleani. Eur. Phys. J. {\bf A2}, 275(1998).
\bibitem{Gla} V.V. Glagolev, JINR Communication E1-12943 (1979).
\bibitem{Shab1} Yu.M. Shabelski. Sov J. Nucl. Phys. {\bf 52}, 984 (1990).
\bibitem{Alk1a} G.A. Korolev et al., Preprint PNPI-2810, St. Petersburg (2009).
\bibitem{Tani3} I. Tanihata, Nucl. Phys. {\bf A488}, 113c (1988).
\bibitem{Tani4} I. Tanihata, D. Hirata, and H. Toki, Nucl. Phys. {\bf A583}, 769c (1995).
\bibitem{Sato} H.Sato and Y. Okuhara, Phys. Lett. {\bf B162}, 217 (1985).
\bibitem{BTR} B.A. Brown, S. Typel, and W.A. Richter, Phys. Rev. {\bf C65},  014612 (2001).
\bibitem{HanJon} P.G. Hansen and B. Jonson, Europhys. Lett. {\bf 4} 409 (1987).
\bibitem{Tani6} I. Tanihata et al., Phys. Lett. {\bf B287}, 307 (1992).
\bibitem{EHMS} H. Esbensen et al., Phys. Rev. {\bf C76}, 024302 (2007).
\bibitem{Shul} N.B. Shulgina, B. Jonson, and M.V. Zhukov, Nucl. Phys.  {\bf A825}, 175 (2009).
\bibitem{FJR} D.V. Fedorov, A.S. Jensen, and K. Riisager, Phys. Rev. {\bf C50},  2372 (1994).
\bibitem{FJR1} D.V. Fedorov, A.S. Jensen, and K. Riisager, Phys. Rev. Lett. {\bf 73},  2817 (1994).
\bibitem{Efi} V.N. Efimov, Phys. Lett. {\bf B33}, 563 (1970).
\bibitem{YFH} M.T. Yamashita, T. Frederico and M.S. Hussein, Mod. Rev. Lett.  {\bf A21}, 1749 (2006).
\bibitem{Hoyle} F. Hoyle et al., Phys. Rev. {\bf 92}, 1095 (1953).
\bibitem{Hoyle1} F. Hoyle et al., Astrophys. J Suppl. {\bf 1}, 121 (1954).
\bibitem{Livio} M. Livio et al., Nature {\bf 340}, 281 (1989).
\bibitem{Rob} F..Robicheaux, Phys. Rev. {\bf A60}, 1706 (1999).
\bibitem{YTF} M.T. Yamashita, L.Tomio, and T. Frederico, Nucl. Phys. {\bf A735}, 40 (2004).
\bibitem{HJJ} P.G. Hansen, A.S. Jensen, and B. Jonson, Annu. Rev. Nucl. Part. Sci. {\bf 45}, 591 (1995).
\bibitem{ElTa} P.J. Ellis and Y.C. Tang, Phys. Rev. Lett. {\bf 56}, 1309 (1986).
\bibitem{Riis} K. Riisager, Rev. Mod. Phys. {\bf 66}, 1105 (1994).
\bibitem{Tani12} I. Tanihata et al., Phys. Lett. {\bf B289}, 261 (1992).
\bibitem{Ryj} V.L. Ryjkov et al. Phys. Rev. Lett {\bf 101}, 012501 (2008).
\bibitem{Tani5} N. Hukunishi, T. Otsuka, and I. Tanihata, Phys. Rev {\bf C48}, 1648 (1993).
\bibitem{Tani9} I. Tanihata, Prog. Part. Nucl. Phys. {\bf 35}, 505 (1995).
\bibitem{Tani7} I. Tanihata, J. Phys. {\bf G22}, 157 (1996).
\bibitem{Koba} T. Kobayashi et al., Phys. Rev. Lett. {\bf 60}, 2599 (1988).
\bibitem{Khal} J. Al-Khalili and F. Nunes, J. Phys. {\bf G29}, R89 (2003).
\bibitem{Tani10} D. Hirata et al., Phys. Rev. Lett. {\bf 75}, 3241 (1995).
\bibitem{Zhu} M.V. Zhukov et al., Phys. Rept. {\bf 231}, 151 (1993).
\bibitem{Bert} C.A. Bertulani, L.F. Cairo, and M.S. Hussein, Phys. Rept. {\bf 226}, 281 (1993).
\bibitem{ATT} J.S. Al-Khalili, J.A. Tostevin, and FI.J. Tompson, Phys. Rev. {\bf C54}, 1843 (1996).
\bibitem{TA} J.A. Tostevin and J.S. Al-Khalili, Phys. Rev. {\bf C59}, R5 (1999).
\bibitem{AOST} B. Abu-Ibragim et al. Comput. Phys. Commun. {\bf 151}, 369 (2003).
\bibitem{TJA} J.A. Tostevin, R.C. Johnson, and J.S. Al-Khalili, Nucl. Phys.  {\bf A630}, 340c (1998).
\bibitem{Moon} C.-B. Moon et al. Phys. Lett {\bf B297}, 39 (1992).
\bibitem{Take} M. Takechi et al., Phys. Rev. {\bf C79}, 061601 (2009).
\bibitem{Shuk} P. Shukla, Phys. Rev. {\bf C67}, 054607 (2003).
\bibitem{DDP} N.J. DiGiacomo, R.M. DeVries, and J.C. Peng, Phys. Rev. Lett. {\bf 45}, 527 (1980).
\bibitem{Kox} S. Kox et al., Phys. Rev. {\bf C35}, 1678 (1987).
\bibitem{Tani8} I. Tanihata, Prog. Part. Nucl. Phys. {\bf 35}, 505 (1995).
\bibitem{You} B.M. Young et al. Phys. Rev. Lett {\bf 71}, 4124 (1993).
\bibitem{Smi} M. Smith et al. Phys. Rev. Lett {\bf 101}, 202501 (2008).
\bibitem{Mitt} W. Mittig et al. Phys. Rev. Lett {\bf 59}, 1889 (1987).
\bibitem{Obu} M.M. Obuti et al., Nucl. Phys. {\bf A609}, 74 (1996).
\bibitem{Mina} T. Minamisono et al. Phys. Rev. Lett {\bf 60}, 2599 (1988).
\bibitem{Oza3} A. Ozawa et al., Phys. Lett. {\bf B334}, 18 (1994).
\bibitem{Tani11} I. Tanihata, Phys. Lett. {\bf B206}, 592 (1988).
\bibitem{Oza4} A. Ozawa et al., Preprint RIKEN-AF-NP-294 (1998).
\bibitem{Chul} L. Chulkov et al., Nucl. Phys. {\bf A603}, 219 (1996).
\bibitem{Bert1} C.A. Bertulani, in 11th Int. Conference on Nuclear Reaction Mechanisms, Varenna (Italy), Villa Monastero, June 12 - 16, 2006; arXiv:nucl-th/0607024v1.
\bibitem{Wan1} Wang Zaijun and Ren Zhongzhou, Science in China, ser. G {\bf 47}, 42 (2004).
\bibitem{Dra} G.W.F. Drake and Z.-C. Yan, Nucl. Phys. {\bf A790}, 151c (2007).
\bibitem{PuPa} M. Puchalski and K. Pachucki, Phys. Rev. {\bf A78}, 052511 (2008).
\bibitem{BRi} E. Borie and G.A. Rinker, Phys. Rev. {\bf A18}, 324 (1978).
\bibitem{Wan} L.-B. Wang et al., Phys. Rev. Lett. {\bf 93}, 142501 (2004).
\bibitem{Mue} P. Mueller et al., Phys. Rev. Lett. {\bf 99}, 152502 (2007).
\bibitem{Ewa} G. Ewald et al., Phys. Rev. Lett. {\bf 93}, 113002 (2004).
\bibitem{Busha} B.A. Bushaw et al., Phys. Rev. Lett. {\bf 91}, 043004 (2003).
\bibitem{DeJa} C.W. De Jager, H. De Vries, and C. De Vries, Atom. Data Nucl. Data Tables {\bf 14}, 479 (1974).
\bibitem{FCN} B.A. Bushaw et al., Phys. Rev. {\bf C79}, 021303 (2009).
\bibitem{San} R. Sanchez et al., Phys. Rev. Lett. {\bf 96}, 033002 (2006).
\bibitem{Nor} W. N\"ortersh\"auer et al., Phys. Rev. Lett. {\bf 102}, 062503 (2009).
\bibitem{Shimo} S. Shimoura et al., Phys. Lett. {\bf B348}, (1995) 29.
\bibitem{Alk1} G.D. Alkhazov, A.V. Dobrovolsky, A.A. Lobodenko, Yad. Fiz. {\bf 69} (2006) 1, Phys. At. Nucl. {\bf 69}, (2006) 1124.
\bibitem{Bertul} C.A. Bertulani and M.S. Mussein, Phys. Rev. {\bf C76}, 051602(R) (2007).
\bibitem{Ikeda} K.Ikeda et al., Nucl. Phys. {\bf A722}, 335c (2003).
\bibitem{Suzu} Y. Suzuki and  Y. Tosaka, Nucl. Phys. {\bf 517} (1990) 599.
\bibitem{Aumann} T. Aumann et al., Phys. Rev. {\bf C59}, (1999) 1252.
\bibitem{Alk2} G.D. Alkhazov, Yad. Fiz. {\bf 63} (2000) 285, Phys. At. Nucl. {\bf 63} (2000) 229.  
\bibitem{Naka1} T. Nakamura et al., Phys. Rev. Lett. {\bf 96}, 252502 (2006).
\bibitem{Naka2} T. Nakamura et al., Phys. Lett. {\bf B331}, (1994) 296.
\bibitem{Hagi} K. Hagino and H. Sagawa, Phys. Rev. {\bf C76} (2007) 047302.
\bibitem{Alk-Lob1} G.D. Alkhazov and A.A. Lobodenko, Pis'ma Zh. Eksp. Teor. Fiz. {\bf 55}, 377 (1992), JETP Lett. {\bf 55}, 379 (1992).
\bibitem{Alk-Lob2} G.D. Alkhazov and A.A. Lobodenko, Proc. Int. Conf. on Nuclei far from Stability, Bernkastel-Kues, Germany, IOP Publishing, Inst. Phys. Conf. Ser. {\bf 132} Sect. 3, 341 (1992).
\bibitem{Alk-Lob3} G.D. Alkhazov and A.A. Lobodenko, Yad. Fiz. {\bf 56}, 89 (1993), Phys. At. Nucl. {\bf 56}, 337 (1993).
\bibitem{Neum} S.R.Neumaier et al., Nucl. Phys. {\bf A712}, 247 (2002).
\bibitem{Dobr} A.V. Dobrovolsky et al., Nucl. Phys. {\bf A766}, 1 (2006).
\bibitem{Vorob} A.A. Vorobyov et al., Nucl. Instr. Meth. {\bf 119}, 509, 1974.
\bibitem{Eldysh} Yu.N. Eldyshev, V.N. Lukyanov, Yu.S. Pol, Yad. Fiz. {\bf 16}, 506 (1972), Sov. J. Nucl. Phys. {\bf 16} 282 (1973).
\bibitem{Al-Kha} J.S. Al-Khalili and J.A. Tostevin, Phys. Ev. {\bf C57} 1846 (1998).
\bibitem{Kis} O.A. Kiselev et al., Eur. Phys. J {\bf A25}, 215 (2005).
\bibitem{Bert2} C.A. Bertulani, 	arXiv:0908.3275v2 [nucl-th].
\bibitem{GANIL1} W. Mittig, J. Phys. G: {\bf 24} (1998) 1331.
\bibitem{GANIL2} B. Jackuot et al., arXiv:nucl-ex/0502016v1
\bibitem{NuSTAR1} R. Kr\"ucken, J. Phys. G: {\bf 31} (2005) S1807.
\bibitem{NuSTAR2} N. Kalantar-Nayestanaki, Acta Phys. Pol. {\bf 41} (2010) 481.
\bibitem{NuSTAR3} H. Simon, Nucl. Phys. {\bf A 787} (2007) 102.
\bibitem{Riken1} Y. Yano, Nucl. Instr. Meth. {\bf B 261} (2007) 1009.
\bibitem{Riken2} H. Sakurai, Eur. Phys. J. Special Topics {\bf 150} (2007) 249.
\bibitem{FRIB1} M. Thoennessen, Nucl. Phys. {\bf A 834} (2010) 688c.
\bibitem{FRIB2} http://www.frib.msu.edu.

\end{thebibliography}
\end{document}